\documentclass[sigconf, nonacm]{acmart}

\usepackage{soul}
\usepackage{xcolor} 
\usepackage{listings}

\definecolor{diffred}{rgb}{0.8,0.0,0.0}
\definecolor{diffgreen}{rgb}{0.0,0.5,0.0}
\setstcolor{diffred}

\AtBeginDocument{%
  }

\setcopyright{none}
\copyrightyear{2018}
\acmYear{2018}
\acmDOI{XXXXXXX.XXXXXXX}

\acmConference[Conference acronym 'XX]{Make sure to enter the correct
  conference title from your rights confirmation emai}{June 03--05,
  2018}{Woodstock, NY}
\acmISBN{978-1-4503-XXXX-X/18/06}

\usepackage{xspace}

\newcommand{\experiment}{\textsc{SemanticCommit}\xspace}
\newcommand{\control}{\textsc{Canvas}\xspace}
\newcommand{\sayit}[1]{``\textit{#1}''}

\newcommand{\minipar}[1]{\textbf{#1}}
\newcommand{\boldsec}[1]{\subsubsection{\textbf{#1}}}

\usepackage{graphicx}
\usepackage{subcaption}
\usepackage{url}
\usepackage{enumitem}

\setlist[itemize]{left=4pt,topsep=3pt}
\setlist[enumerate]{left=4pt,topsep=3pt}

\begin{document}

\title{Semantic Commit: Helping Users Update Intent Specifications for AI Memory at Scale}

\author{Priyan Vaithilingam\textsuperscript{1}, Munyeong Kim\textsuperscript{2}, Frida-Cecilia Acosta-Parenteau\textsuperscript{2}, Daniel Lee\textsuperscript{3}, Amine Mhedhbi\textsuperscript{4}, Elena L. Glassman\textsuperscript{1}, Ian Arawjo\textsuperscript{2}}

\thanks{\textsuperscript{1}Harvard University, Boston, Massachusetts, USA. \texttt{\{pvaithilingam, glassman\}@g.harvard.edu}}
\thanks{\textsuperscript{2}Montréal HCI, Université de Montréal, Montréal, Québec, Canada. \texttt{\{kim.munyeong, frida-cecilia.acosta-parenteau, ian.arawjo\}@umontreal.ca}}
\thanks{\textsuperscript{3}Adobe Inc., San Jose, California, USA. \texttt{dlee1@adobe.com}}
\thanks{\textsuperscript{4}Data Systems Group, Polytechnique Montréal, Montréal, Québec, Canada. \texttt{amine.mhedhbi@polymtl.ca}}

\renewcommand{\shortauthors}{Vaithilingam et al.}

\begin{abstract}
  How do we update AI memory of user intent as intent changes? We consider how an AI interface may assist the integration of new information into a repository of natural language data. Inspired by software engineering concepts like impact analysis, we develop methods and a UI for managing semantic changes with non-local effects, which we call ``semantic conflict resolution.'' The user commits new intent to a project---makes a ``semantic commit''---and the AI helps the user detect and resolve semantic conflicts within a store of existing information representing their intent (an ``intent specification''). We develop an interface, \textsc{SemanticCommit}, to better understand how users resolve conflicts when updating intent specifications such as Cursor Rules and game design documents. 
  A knowledge graph-based RAG pipeline drives conflict detection, while LLMs assist in suggesting resolutions. We evaluate our technique on an initial benchmark. Then, we report a 12 user within-subjects study of \textsc{SemanticCommit} for two task domains---game design documents, and AI agent memory in the style of ChatGPT memories---where users integrated new information into an existing list. Half of our participants adopted a workflow of \textit{impact analysis}, where they would first flag conflicts without AI revisions then resolve conflicts locally, despite having access to a global revision feature. We argue that AI agent interfaces, such as software IDEs like Cursor and Windsurf, should provide affordances for impact analysis and help users validate AI retrieval independently from generation. Our work speaks to how AI agent designers should think about updating memory as a process that involves human feedback and decision-making.

\end{abstract}



\definecolor{grayborder}{RGB}{200, 200, 200}

\begin{teaserfigure}
  \centering
  \includegraphics[width=\textwidth]{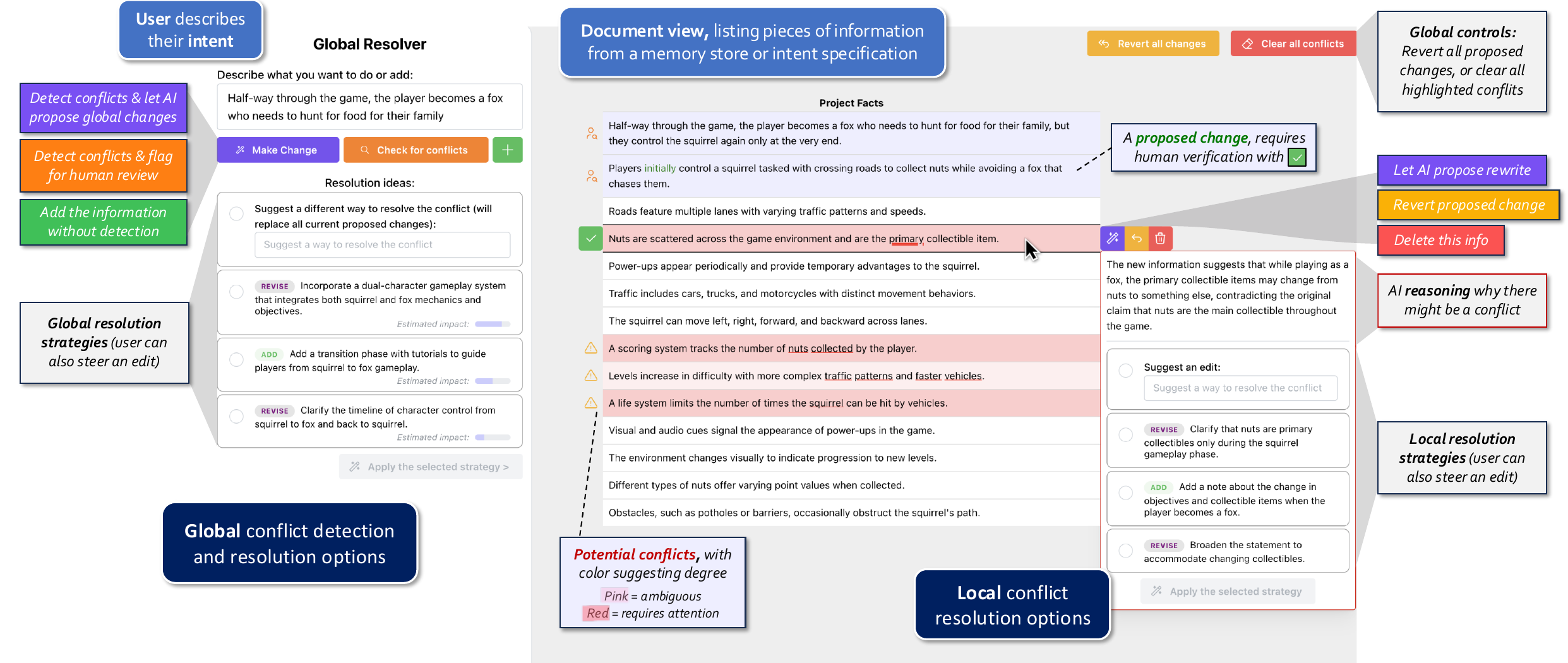}
  \caption{Our \textsc{SemanticCommit} interface, providing users myriad ways to detect and resolve conflicts at global and local levels. Our prototype was used as a probe to better understand the needs of users for integrating new information into lists of prior information akin to AI agent memory or requirements lists. The screenshot depicts a short list describing a ``Squirrel Game,'' where the user is integrating a new feature. Potential conflicts are highlighted in red and pink to mark degree, and the AI has added a new piece of information to the store and proposed an edit to another piece, both marked for human verification.}
  \label{fig:teaser}
\end{teaserfigure}

\received{20 February 2007}
\received[revised]{12 March 2009}
\received[accepted]{5 June 2009}

\settopmatter{printfolios=true}

\maketitle

\clearpage

\section{Introduction}

In the near-future, people may coordinate with AI agent systems through project-specific documents that represent accumulations of user intent \cite{vaithilingam2024imagining, zamfirescu2025beyond, ma2024rope}---lists that we call \textbf{intent specifications}. These human-readable accumulations of design requirements, user goals and preferences reify common ground~\cite{clark1991grounding, bansal2024challenges, vaithilingam2024imagining} between humans and AI systems, grounding AI decision-making by keeping track of details and goals, surfacing implicit assumptions made by AI, and acting as a intermediate representation of an AI system's `understanding' which the user can inspect and edit (Figure~\ref{fig:intent-specs}).

We dream of a world in which people can make \textbf{semantic commits}: committing ideas and details to projects like they commit code, and dealing with the ``merge conflicts'' that may occur. One key challenge standing in the way of this paradigm shift is \textbf{\emph{integration}}: how to responsibly and verifiably integrate new information into a repository of natural language \cite{vaithilingam2024imagining}, e.g., to update an AI agent's memory of user intent in a reviewable, concise, and accurate manner, such that the memory remains aligned. How can technology assist with the integration of a new piece of information into an existing repository \textit{at scale} (e.g., a design document, a requirements list, documentation, a wiki, novel, etc.)? The new information may conflict with prior information---something may become \emph{inconsistent} or \emph{contradictory}. Changing existing information can incur the same effect. We frame this challenge for the community as \textbf{\textit{semantic conflict detection}} and \textbf{\textit{semantic conflict resolution}}, since it operates at the level of \textit{semantics} and \textit{concepts}, unlike past techniques in software engineering and document synchronization that operate on pre-defined structure and syntax.

Over brief time-frames and short documents, simple methods---such as using LLMs to regenerate entire documents or apply string replace operations \cite{Labin2024Beyond}---can perform edits, but as humans interact with agents over long time-frames and complex projects, these methods cease to function at scale. Simple vector store %
architectures, %
seen in retrieval-augmented generation (RAG), also face challenges, since detecting semantic conflicts frequently requires multi-hop reasoning, a well-known failure mode \cite{gutierrez2024hipporag, graphrag}. \textit{How} to resolve a conflict is also often subjective \cite{jiang2022investigating, chen2025onreference}, and therefore a problem for HCI, as for example, particular conflict resolutions may incur \textit{cascades} where solving one problem creates another. Systems thus need ways not only of \textit{identifying} conflicts and inconsistencies efficiently and accurately at scale, but of \textit{interactively assisting users in conflict resolution} in a way that a) helps users reflect and b) foresee the impact of changes, c) only makes the necessary changes without touching other information, and d) minimizes user effort while maximizing changes' alignment with user intent. Downstream AI systems could use conflict detection results to, e.g., decide whether to perform grounding acts~\cite{shaikh2025navigatingriftshumanllmgrounding, shaikh2024grounding} such as request for clarification. %

To help researchers better understand the problem of updating AI memory of user intent in an aligned manner, in this paper, we provide several contributions to the literature. %
We:

\begin{enumerate}
    \item Define the term \textit{intent specification} to name grounding documents that coordinate with AI agents, such as user-defined ``memory'' lists for Claude Code~\cite{claude_code_overview}.
    \item Provide design goals for AI-assisted interfaces for semantic conflict detection and resolution, inspired by related literature such as impact analysis in software engineering. %
    \item Develop an interface, \textsc{SemanticCommit}, iterating its design over two pilot studies. Our system implements a range of affordances for conflict detection and resolution and is intended as a probe of user behavior.
    \item Introduce an architecture for semantic conflict detection at scale. Our approach uses induced knowledge graphs, adapting emerging architecture in retrieval-augmented generation (RAG)~\cite{gutierrez2024hipporag}. To test and compare our approach to prior approaches in the literature, we also provide a custom benchmark on intent specifications across several settings. %
    \item Provide empirical insights from a within-subjects user study, examining how users detect, understand, and resolve conflicts when updating intent specifications for AI memory, comparing \textsc{SemanticCommit} to OpenAI's ChatGPT Canvas. %
\end{enumerate}

Our findings suggest that AI agent interfaces should  enable users to perform \textbf{\textit{impact analysis}}, separating retrieval from generation---steps that are currently conflated in many AI-powered software engineering IDEs. Surprisingly, although users appeared more engaged when using \textsc{SemanticCommit}, they did not report significantly higher workload than the more automated Canvas UI. This suggests that \textbf{the benefits of improved control can offset the cost of manual review,} possibly by shifting user workload away from metacognitive demands~\cite{tankelevitch2024metacognitive} that users face when prompting in open-ended chat, towards the demands of the actual task, such as reviewing conflicts. %

\section{Motivation: Intent Specifications Ground Human Coordination with AI Agents} \label{motivation}

Humans are increasingly managing and validating the outputs of AI systems that implement entire software systems like games, websites, and apps. To reduce risk and align AI decision-making in user preferences, human-readable documents are emerging as a mechanism to create and maintain common ground~\cite{clark1991grounding} between humans and AI systems acting on their behalf. %

\begin{figure}
  \centering
  \includegraphics[width=\linewidth]{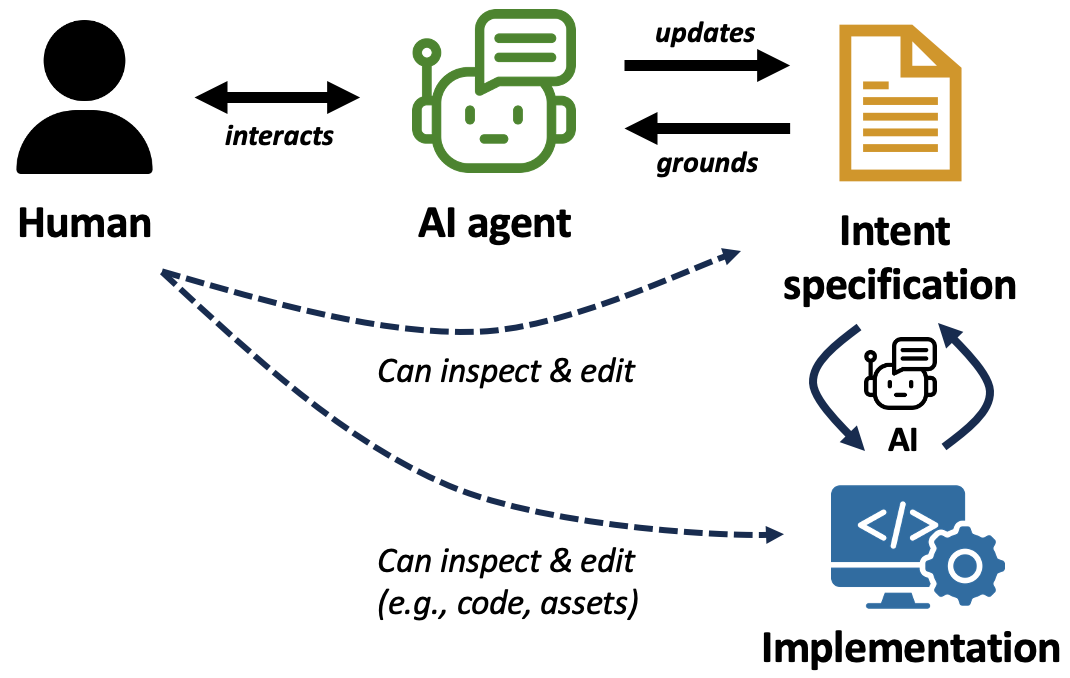}
  \caption{A high-level depiction of our envisioned interaction between humans and AI assistants for long-term projects. The human-readable \textit{intent specification} serves as an intermediate layer for enhancing common ground between the human and the AI, and grounds the AI's decision-making. We assume future AI agents will have a similar intent specification layer. Our project  squarely concerns how the AI \textit{updates} this memory in a robust, verifiable manner, and in the process might surface conflicts to the user to get their feedback in resolving them.}
  \label{fig:intent-specs}
\end{figure}

Numerous examples are emerging of this interaction paradigm. The AI-powered programming IDE Cursor, for instance, can ground its behavior in user-made ``cursor rules''---markdown documents that AIs read to ground their behavior in user preferences---at both project-specific and global levels \cite{cursor_rules_for_ai}.\footnote{People have started crowd-source these rules: the ``Awesome CursorRules'' repository and CursorList.com include hundreds of rules lists, contributed by everyday users, indexed by programming language, libraries, and use cases. See, e.g., \url{https://github.com/PatrickJS/awesome-cursorrules}.} %
Rules range from sweeping commands, like ``never use apologies,'' to the highly particular, like ``use vectorized operations in pandas and numpy for improved performance.'' Users develop these rules over time across many interactions. Anthropic has also adopted this paradigm: with the Claude Code agent, users  create \texttt{CLAUDE.md} files listing project- and global-level directives; Anthropic's own ``memory best practices'' tell users to format memories as ``bullet points'' and reminds them to manually ``update memories as your project evolves'' %
\cite{claude_code_overview}. These ``memories'' help Claude Code ``remember project conventions, architecture decisions, or coding standards that we want to reference across sessions'' \cite{catwu2025}. %
Not to be outdone, the CEO of Windsurf---a competitor to Cursor---just announced a yet-to-be-implemented ``auto-generated memories'' feature where these memories of user intent are automatically updated by an AI, which will inevitably encounter the very challenges we discuss here.\footnote{https://x.com/vitrupo/status/1900146068030914740}

Everyday users are also increasingly coordinating with AI systems through lists of requirements expressed in natural language. For instance, users are generating games from specs that resemble lists of software requirements. Here is an excerpt from a real user,\footnote{\url{https://github.com/vnglst/when-ai-fails/blob/main/shepards-dog/README.md}} to give readers a sense of how these rules appear in practice: 

\begin{itemize}
    \item The dog barks when the player clicks or taps on the screen, making the sheep move faster
    \item Sheep should react realistically to the dog's presence %
    \item When frightened, the flock should scatter
\end{itemize}

This user's example, which in total has 27 requirements, is only the \textit{start} of an interaction with an AI agent. As the user interacts and projects grow in complexity, future AI systems will need to assist in the extension and updating of these rules and details. %

We call these lists---cursorrules, \texttt{CLAUDE.md} files, user directives, AI memory of user intent, etc.---\textbf{intent specifications}, adapting and broadening the notion of \textit{requirement specifications} in software engineering. Intent specifications are \textbf{\textit{evolving, comprehensible documents of user intent that ground AI decision-making and reify common ground between humans and AI systems}}. We introduce \textit{intent specification} to underscore that such documents may not only cover design details or software requirements, but how the AI should communicate to the user, who the user is, the user's goals and dreams, etc. Said differently, an intent specification is not only a description of user intent, but may also include information that helps an AI agent \textit{assume} user intent---i.e., background, assumed preferences---accelerating the establishment of common ground. %
However, unlike a general memory store---which could be an extensive collection of all interactions---intent specifications' purpose is to be reviewable, comprehensible and digestible, to be inspected and edited by humans. In response to edits, the AI will adjust its behavior, such as revising an implementation; the AI may also amend the specification in response to the user or to better reflect new implementation details and assumptions~\cite{zamfirescu2025beyond, vaithilingam2024imagining} (Fig.~\ref{fig:intent-specs}).

As we mention in our introduction, the \textit{integration} of new information into an intent specification is not (always) straightforward. People and ideas change. New information may conflict with prior information, especially as projects and user interactions stretch from days to months and years. Proposed approaches to memory with RAG architectures, which store all memories verbatim, do not account for these potential conflicts (e.g., \cite{LangMem, park2023generative, hou2024my}).%

Our chief insight here that integration of new information is a \textit{process} that can require \textit{interactive}, human-in-the-loop feedback for aligned resolution. Both users and AI systems need support for semantic conflict \textit{detection}---to understand when a conflict has taken place, with what information, and how---as well as \textit{resolution}, as resolving conflicts could involve the revision, addition, or deletion of existing information in a manner that may add or change details. To resolve conflicts, practical assistance may require \textit{clarification of ambiguities}, \textit{constructive negotiation of ideas} \cite{vaithilingam2024imagining}, or \textit{delegation of tasks} \cite{shao2024collaborative, zamfirescu2025beyond}. However, it remains unclear how users update, and want to update, intent specifications in practice. \textbf{What affordances should AI memory interfaces have for the process of integration? How  do users think about semantic conflicts and what needs do they have for resolving them with confidence? How can we help users update intent specifications like \texttt{CLAUDE.md} files with confidence?} %

Before returning to these questions with our system design and user study, we first connect to existing literature that can help shed light on this emerging paradigm. %

\subsection{Related Work} \label{related-work}

\subsubsection{\textbf{Design documents to coordinate work in human teams.}} 
The rise of intent specifications %
mirrors what human teams already do to coordinate actions. 
Across many domains---from product design, to game development, software engineering, UX design, construction, and animation---people standardize the vision (look, feel, goals, plans, etc) of a project in documents that are often called ``design documents'' \cite{boy1997active}.
These documents serve to establish and maintain \textit{common ground} between parties \cite{clark1991grounding}, ensuring each member's independent actions remain \textit{grounded} in shared understanding and objectives. %
In animation, the design document takes the form of model sheets \cite{musburger2017animation}, which standardize how to draw characters and other assets. Game developers use ``game design documents'' (GDDs) to keep development grounded across a team~\cite{colby2019game}. %
In software engineering (SE) and UX research, need-finding studies and client discussions produce a ``system requirements specification'' that is passed off to the software team \cite{loeffler2013developing, hartson2012ux}. Programmers develop ``coding style guides,'' or norms around naming conventions, comments, and writing tests, as well as ``contributing guidelines'' that establish expectations and rules for external contributors.  
In all cases these documents serve to externalize, standardize, and coordinate the high-level \textit{intent} of a team---its objectives, details, procedures, and feel---and are revised as the project proceeds~\cite{colby2019game}. 

Intent specifications, while less formal than code, are a lot like software: they encode dependencies among ideas that constrains future evolution. These dependencies may be “sequential” (i.e., a custom term is defined then used later on) or heterarchical. As projects continue, teams ``commit'' new information to the intent specification, and must resolve outdated or inconsistent dependencies. Maintaining consistency is paramount, because the very purpose of these documents is \textit{to enforce consistency and define standards}. For instance, from a study of game designers:
    \emph{``[She] writes the GDD as she is designing the game... %
    taking anything out of... the GDD that \textbf{conflict} with the \textbf{consistency} of her plot. [She]... \textbf{wrote her entire GDD... as a list}, which she frequently \textbf{added} and \textbf{deleted} from as she designed the game''} \cite[p.9]{colby2019game}. 
The field of requirements engineering in SE also stresses the importance of clarity, conciseness, completeness, and unambiguous requirements~\cite{dermeval2016applications}, with ``commission (inclusion of irrelevant or incorrect details) and omission (exclusion of necessary details)'' additional concerns \cite{ma2024rope}.

Currently, for updating \textit{unstructured} documents, there is no way for those committing changes to see the \textit{semantic} ramifications of their changes on the rest of the document. An example is editing a scene in a novel: would changing a lunch between two characters to a dinner setting impact something hundreds of pages later? These are \textbf{semantic conflicts} that must be \textit{detected} and \textit{resolved} to proceed with confidence. In software engineering, visibility on the ramifications of a feature change or addition is called \textbf{impact analysis}: identifying what parts of the shared context (code repository) will need to be amended, for the change to occur \cite{arnold1996software}. Note that impact analysis is a \emph{sense-making} task, less a coding one: it is more about estimating time and resources required to perform a change, than it is about helping the team actually perform it. %

\subsubsection{\textbf{Human-AI Collaboration Grounded in Shared, Intermediate Representations.}}

HCI has, in a sense, always been about communicating to machines through shared representations \cite{arawjo2020write, heer2015predictive}. However, past shared representations had to be strictly formalized---into programming languages, domain-specific languages (DSLs), schema, etc.---to ensure deterministic outcomes. These shared representations helped negotiate agency between humans and machines \cite{heer2019agency}, but ultimately could only go so far, as end-users always resigned some agency to the representation designer(s) \cite{li2023beyond, vaithilingam2024imagining}. 

Today with LLMs, we are less limited by this constraint, and solutions to the problem of human-machine communication might be better found in cybernetics theory~\cite{beer2002cybernetics} than static formalism. Effective human-AI communication relies upon tight feedback loops~\cite{zamfirescu2025beyond}, but also offering humans \textit{control} in the form of transparency over AI understanding and context. %
Along these lines, emerging HCI research envisions that AI systems will be grounded by shared representations of a more informal nature---lists of directives expressed in natural language \cite{ma2024rope, zamfirescu2025beyond, vaithilingam2024imagining}. Some researchers even argue that these informal expressions of intention will be ``all you need'' \cite{sarkar2024intention, robinson2024requirements}. For instance, Vaithilingam et al. imagine a hypothetical AI game design assistant where the AI ``[integrates user] choices into the project plan'' \cite{vaithilingam2024imagining}, while Zamfirescu et al. explore an iterative design loop with an AI agent that ``tracks decisions that the human has made'' and ``surfaces decisions the LLM has implemented in the code'' in a running list~\cite{zamfirescu2025beyond}. Ma et al. define ``requirement-oriented prompt engineering,'' helping users generate a ``clear, complete requirements'' list prior to prompting an AI to implement software. They stress that making a good list requires skill and support \cite{ma2024rope}. These projects speak to the need for targeted support for updating intent specifications that ground AI behavior. %
Ensuring alignment with user intent (e.g., by reducing inconsistencies) is critical: miscommunications are the chief reason for breakdowns with AI agents \cite{shao2024collaborative}, and the probability of failure compounds as communication continues without addressing misunderstandings \cite{shaikh2025navigatingriftshumanllmgrounding}.

\subsubsection{\textbf{Conflict detection and resolution techniques and interfaces.}} Conflict detection and resolution are classic problems in computing, usually arising in contexts of collaborative information processing to merge asynchronous changes. Engineers have developed techniques such as version control~\cite{fan2012supporting, nelson2019life}, groupware platforms and database synchronization~\cite{phatak1999conflict, litt2022peritext, klokmose2024mywebstrates}, and concurrency-control systems~\cite{fan2012supporting, haibin2005conflict}. %
The `git` command-line interface \cite{Git2025}, for instance, is a popular  version control system where users make ``commits''---a change to a file repository, alongside a pithy message---to keep track of changes. %
Conflict management systems now extend to nearly all 
collaborative computing settings, such as document editing with Conflict-Free Replicated Data Types (CRDTs) \cite{litt2022peritext, klokmose2024mywebstrates}. %

Many text interfaces that support conflict detection and resolution visualize differences between versions as ``diffs'' \cite{hunt1976algorithm}. %
Diff visualizations usually {\textcolor{diffred}{\st{cross out deleted information in red}}} {\textcolor{diffgreen}{\ul{and showcase changes or additions in green underline}}}, a colored version of the method originally proposed by Neuwirth et al.~\cite{neuwirth1992flexible}. The visualization has been extended and applied in other ways~\cite{nugroho2020different, rostami2024gawd}; for instance, when merging branches of a code repository, programming IDEs  show ``merge conflicts'' in-line as highlighted segments of code. %
Recently, Ink\&Switch's Patchwork project~\cite{litt2024universal} explores generalizations of the diff, including summary and diagram diffs. %

Diffs are now being used to showcase edits made by AI. Numerous tools have recently been developed to enhance writing workflows with LLMs, spanning various areas such as story writing~\cite{Chung2022TaleBrush, Clark2018Creative, Yuan2022Wordcraft}, screenplay writing~\cite{Mirowski2023Co-Writing}, poetry~\cite{ghazvininejad2017hafez}, dictation~\cite{lin2024rambler}, and argumentative~\cite{Wambsganss2020AL, Zhang2023VISAR} and scientific~\cite{feng2024cocoa} writing. %
\textsc{InkSync}~\cite{Labin2024Beyond}, for example, is a prototype for executable and  verifiable text editing with LLMs, which shows LLM edits as diffs on the document. To make diffs, \textsc{InkSync} uses string-matching: it relies upon the LLM to reproduce extracts of text to change, and then specify the replacement text; this method is also used by Anthropic Artifacts~\cite{reddit_claude_system_prompt}.

However, there is a limit to which this past work can inform our situation. The above situations and techniques \textit{assume the updated document}. That is, the updated document is a given, and the system needs to detect changes and visualize them to the user. 

When making semantic commits, \textbf{\textit{we do not have, or cannot assume, the updated document as a ground truth}}. Semantic conflict resolution interfaces must therefore go further---not just visualizing what changes \textit{were} made, but what changes \textit{could be} made, \textit{where} they should be made, and \textit{what the effects} might be. Where \textit{diff} interfaces provide feed\textit{back}, semantic diff interfaces must offer feed\textit{forward}: affordances that help the user foresee what the result of an action may be \cite{vermeulen2013feedforward, min2025feedforward}. This new situation resembles ``impact analysis'' in software engineering %
\cite{figueiredo2012wolf_impactanalysis}, but here we cannot assume well-structured data like a programming language.

\subsubsection{\textbf{Natural Language Inference, Reference Ambiguity, and Knowledge Graphs}} \label{rw-nli}

Finally, the technical side of our work relates to natural language inference (NLI), a research area in NLP~\cite{jiang2022investigating} that concerns the classification task: Given two sentences---a premise sentence and a hypothesis--does the hypothesis sentence follow from (\textit{entailment}), \textit{contradict}, or bear a \textit{neutral} relationship to the premise? HCI scholars have applied NLI to data annotation~\cite{wang2024human}, in-situ summaries~\cite{liu2024selenite}, and LLM response consistency~\cite{cheng2024relic}. Our discussion of NLI provides additional context for our system design. %

Detecting conflicts is by no means an objective task; human annotators frequently disagree~\cite{jiang2022investigating, chen2025onreference}. Jiang \& de Marneffe investigated reasons for human disagreement during NLI classification~\cite{jiang2022investigating} and argue for a fourth category, ``complicated,'' which increased model recall. Their goal was ``not necessarily to maximize accuracy. A model that can recall the possible interpretations is preferred to a model that misses them'' \cite[p. 1365]{jiang2022investigating}. Chen et al.~\cite{chen2025onreference} also introduced a fourth category, ``ambiguous,'' to denote situations where ``it is unclear whether the claim and the evidence refer to the same context... [i.e.,] there exist multiple possible assignments or interpretations.'' The authors refer to this as \textbf{reference ambiguity}---when the two sentences \textit{could} coexist, but it is unclear---and found that it explained many annotator disagreements~\cite{chen2025onreference}. 

NLI appears in recent discussions on the future of SE, which propose that LLMs may be used for formal requirements analysis \cite{arora2024advancing, terragni2025futuresoftwareengineering}. A few benchmarking studies test this hypothesis; e.g., Lubos et al.~\cite{lubos2024leveraging} studied how LLMs can provide quality feedback on requirements, while Fantechi et al.~\cite{fantechi2023inconsistency} analyze an LLM's ability to detect inconsistencies. Importantly, Fantechi et al.'s method simply fed in the entire list into the LLM and asked it to detect conflicts; they found that LLMs could only process ``short requirement documents'' this way, with performance falling off quickly for longer ones. They conclude that despite lower accuracy compared to humans, ``manual detection of inconsistencies is more expensive,'' growing quadratically with list size, ``whereas examining [LLM] answers to distinguish true from false positives is a much lighter task'' \cite[p. 338]{fantechi2023inconsistency}. Fazelnia et al.~\cite{fazelnia2024lessons} also trained an NLI model to analyze requirements lists, and concluded that %
NLI models suffered specifically in multi-hop conflict detection. %

To better capture dependencies among requirements, researchers studying requirements engineering in SE proposed ontology extraction, where a system generates a knowledge graph (alternatively called a web ontology~\cite{antoniou2009web}) with nodes for entities and edges for relationships between them. For instance, Hsieh et al.~\cite{HSIEH2011288} extract a domain-specific ontology by mining information from textbooks (analogous to inductive coding). %
Most relevant is research that explores mapping formal software requirements as knowledge graphs~\cite{dermeval2016applications}, a method introduced by Kaiya and Saeki~\cite{kaiya2006using}. The authors use a web ontology as a visualization technique to help SWEs in writing more “correct,” “complete,” “consistent,” and “unambiguous” software requirements~\cite{dermeval2016applications}. %
Such graph-based visualizations have also supported impact analysis; for instance, \textit{Wolf}~\cite{figueiredo2012wolf_impactanalysis} shows the impact of proposed changes via a dependency graph. This work informed our decision to use knowledge graphs (Section~\ref{tech-eval}).

\section{Design Goals for Interfaces for Semantic Conflict Detection and Resolution}

Here we chronicle our initial design goals for \textsc{SemanticCommit}, as well as our revised goals as the result of two pilot studies. 

We wanted to design a prototype to better understand what interface affordances users need to facilitate robust and trustworthy updates to intent specifications in a manner that 1) maintained their alignment with user intent and 2) kept unrelated information untouched. We thus went for a kitchen-sink approach: to include a variety of features that users may, or may not, choose to engage in, that seemed reasonable based on past conflict resolution interfaces, and observe what features users find most important and how they use these features in concert. Based on our review of past literature on conflict detection, resolution, and AI-assisted writing, we identified an initial set of design requirements for \textsc{SemanticCommit}:

\begin{itemize}
    \item \textbf{Foresee impact}: Users should be able to perform \textit{\textbf{semantic impact analysis}}---foresee the potential impact of a change, without actually making any changes~\cite{arnold1996software, figueiredo2012wolf_impactanalysis}.
    \item \textbf{Detect conflicts}: The system should help the user \textit{detect} potential conflicts or contradictions, between existing information and the new information being introduced. 
    \item \textbf{Understand conflicts}: The system should help the user \textit{understand} the reason for conflicts, to reduce cognitive load. 
    \item \textbf{Leave non-conflicting information unchanged}: Integrating new information should only touch pieces of information that are in conflict, and leave others unchanged.
    \item \textbf{Support local changes}: Users should be able to inspect proposed changes in situ and decide whether to accept, reject, or further revise (such as via a ``diff'' view).
    \item \textbf{Assist conflict resolution}: The system should help users resolve conflicts at both global (i.e., entire document) and local levels. The AI should suggest possible \textit{resolution strategies}.
    \item \textbf{Revert changes}: Proposed changes (edits) should be able to be reverted at global and local levels (i.e., to cancel specific revisions or back out from a wide-scale change).
    \item \textbf{Edit manually at any time}: Users should be able to manually edit or add information at any time, should they choose.
    \item \textbf{Work at scale}: The system should work at scale, i.e., for lengthy intent specifications, without introducing latency. The user should not have to worry about document length.
\end{itemize}

Note that there are other design goals which are important to \textit{general} user interfaces for managing AI memory---such as version control, branching, and navigation~(see \textit{Memolet}~\cite{yen2024memolet})---but we do not consider them here.\footnote{In particular, in real-life intent specifications like Cursor Rules, users sometimes group lines together; we chose a simple list to avoid complexity in our initial design.}

It's critical to note that while some design goals overlap with document editing interfaces, \textit{a primary goal of our research is to produce design implications for situations where there may be no manual document view}---e.g., situations where the user is communicating entirely through a chat UI, where the AI is managing the intent specification for them and may surface conflicts in a different, constrained manner (and decide whether, when, and how to do so). We intend that semantic commit will eventually be \textit{a programmatic API} for helping developers update intent specifications that ground AI agent systems. %
Thus, we designed our interface to purposefully constrain editing to separate pieces of information---``memories,'' details or rules---rather than enabling the user to perform freeform writing tasks (i.e., think OpenAI ChatGPT's memory store~\cite{openai_memory_controls_2024}, rather than Microsoft Word).

\subsection{Early Prototype and Pilot Feedback} \label{pilots}

Our explorations went through substantial iterations and prompt prototyping over a period of eight months, evolving in response to two pilot studies and progressing from a card-based interface to a list of texts. We chronicle our early design and formative studies. 

From our design goals, we built an initial prototype, where pieces of information were written on cards akin to post-its and could be freely moved. This interface was limited to prompting our conflict detection feature, and studied how users would integrated changes into (a chunked version of) the game design document for the unpublished LucasArts game Labyrinth~\cite{fox1986labyrinth}. In this early prototype, cards were only marked as either in conflict or not. 

We ran one pilot study with five users of our card-based interface, and a second with four users of a revised interface. Key takeaways:

\begin{itemize}
    \item The color-coding of cards marked as conflicts drew user attention sometimes entirely away from manual inspection of non-marked cards. Possibly in reaction, all pilot users \textbf{preferred higher recall over precision}. They viewed false negatives (missed detections of true conflicts) as catastrophic, while false positives were easily handled with a quick skim.
    \item When asked, participants expressed a preference for a \textbf{structured, sequential document view,} over the cards interface. One reason may be that users became fixated on sorting the cards, another could be that documents are more familiar.\footnote{This preference seems to map to the ``cursorrules''-like situations of editing Markdown documents, which weren't popular at the time of our pilot.}
    \item Users wanted finer-grained insight into the \textbf{degree of conflict}. Users wanted a quick visual way to understand where they should spend their limited attention.
    \item Participants would \textbf{iterate on their prompts} to the conflict detector and resolver, in case the output did not exactly match their intent. It seemed less important that AI sometimes made mistakes, and more that they were easily fixable. 
    \item In post-interviews, users suggested that \textbf{the degree to which they trusted the AI depends on their degree of investment in the information}. If they felt invested, they would trust the AI \textit{less} to make direct changes. %
\end{itemize}

\begin{figure*}
    \centering
    \includegraphics[width=\linewidth]{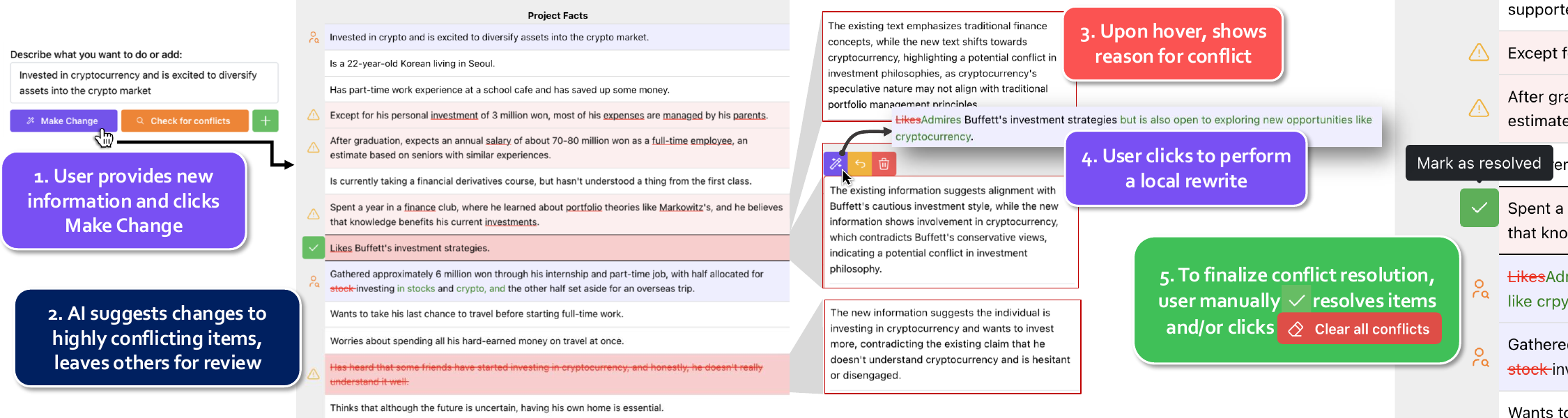}
    \caption{Example of our \textsc{SemanticCommit} workflow, showing one process of integrating new information into an AI memory of the financial habits of a South Korean student. 1. The user has described a new piece of information and pressed Make Change. 2. \textsc{SemanticCommit} detects conflicts and suggests changes to items it deems the most conflicting, leaving other conflicts for human review. 3. The user hovers over conflicting items to view the AI's reasoning. 4. For one item, they click a button to let the AI make a local rewrite. The user can continuing editing, manually revising, reverting suggested changes, or deleting items at will. 5. When they feel done, they manually resolve items and/or clear remaining conflicts with a global action. (Alternatively, the user could have clicked Check for Conflicts to only perform detection, then handled conflicts locally.)}
    \label{fig:workflow}
\end{figure*}

In response, we added more design goals to our initial list:

\begin{itemize}
    \item \textbf{Recall-first}: Favor recall over precision for conflict detection.
    \item \textbf{List view}: The system should prefer more standard document views, which present manageable chunks of information sequentially, than open-ended diagramming canvases. 
    \item \textbf{Visualize degree}: The system should help users understand the degree or importance of a conflict at a glance. 
    \item \textbf{Help user recover from AI errors}~\cite{nielsen1994enhancing}: The system should support fast iteration, in case of AI mistakes, by allowing the user to steer the detector or resolver with a prompt. %
\end{itemize}

Based on these goals and feedback, we adjusted our interface and study protocol. %
The most important change we made was how strict our conflict detection retriever and filtering prompt was: we loosened it considerably, to enhance recall at the expense of precision. We also added a third classification, ``ambiguous,'' to imply a lesser ``degree'' of conflict, a decision solidified after review of  papers in NLI \cite{chen2025onreference, jiang2022investigating}. Ambiguous conflicts appear as a softer pink color to imply reduced importance, directness, or confidence that the information is truly in conflict.\footnote{As we rely upon LLMs, this is not an exact science. Indeed, the aforementioned NLI papers also show that even with human annotators, there is little consistent reason why something is categorized as ``ambiguous'' or ``complicated''~\cite{jiang2022investigating, chen2025onreference}.} %
This prompt engineering was a delicate balance: too restrictive and the system tends to only rarely include ambiguous options; too generic and it flags almost all pieces of information as potentially conflicting. We iterated on our system decision choices with more confidence by validating changes against custom benchmarks, which we discuss in Section~\ref{tech-eval}.

\section{\textsc{SemanticCommit} User Interface}

Here we overview our final design and walkthrough examples of usage.  Figure~\ref{fig:teaser} shows our prototype, with global operations:

\begin{itemize}
    \item \raisebox{-0.5ex}{\includegraphics[height=1.1em]{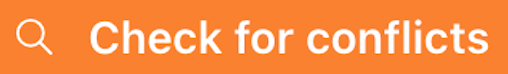}} The %
    Check for Conflicts button provides the ability to perform \textit{impact analysis} \cite{arnold1996software}, which only highlights potential conflicts without suggesting changes, allowing the user to get a sense of how much effort a change might require. They may choose to manually resolve each conflict, or back out and decide upon a different course of action. %
    \item \raisebox{-0.5ex}{\includegraphics[height=1.1em]{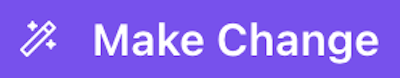}} The Makes Changes button performs Check for Conflicts then lets the AI propose a rewrite. The back-end uses the same procedure as check for conflicts, then performs a global rewrite of all detected conflicts in order to incorporate new information. Critically, the LLM can decide not to rewrite information, even after it has been flagged (this is to avoid redundant changes); flagged conflicts that were not changed remain highlighted for human review. 
    \item \raisebox{-0.5ex}{\includegraphics[height=1.1em]{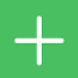}} The Add Info button allows the user to manually add a piece of information. %
\end{itemize}

More features are shown in Fig.~\ref{fig:teaser}. Local conflict resolution options include letting the AI rewrite, steering a rewrite, applying a suggested resolution strategy, reverting a change, and deleting the information. Global conflict resolution options complement this, allowing the user to steer a global rewrite via a prompt or choose a suggested resolution strategy. Users can also perform global actions to Revert All proposed changes, or Clear All Conflicts (putting all pieces of information back into a neutral state). Finally, red {\setulcolor{red}\ul{underlines}} are an experimental feature that suggests words which contributed the most to the conflict (in Fig.~\ref{fig:teaser}, ``{\setulcolor{red}\setul{1.5pt}{1.5pt}\ul{primary}}'' is bold-underlined to imply that nuts are likely no longer the primary collectible when the player is a fox). 

The only feature missing from our figure is a ``request intent clarification'' pop-up that appears when the AI classifies a user request as potentially resulting in many changes (Section~\ref{example-gdd}). We observed that high-impact changes, like changing a game's setting from Mars to Venus, could incur many second-order effects and deserves an additional clarification round before proceeding with  (relatively more costly) conflict detection.\footnote{Our prompt to the AI for this step is simple and more of a prototype: here, we simply feed the entire context in alongside the user's change, and ask the AI to provide a question if it decides the change is high-impact enough to deserve clarification.}

\subsection{Walkthrough of usage}

Let's walk through three examples of system usage in different domains: an investment advisor agent, updating the directives for an AI software engineer, and updating a game design document. 

\subsubsection{\textbf{Updating Memory of an Investment Advice Agent.}} As a simple example, imagine an AI agent for investment advice has accumulated a memory of the user, a South Korean college student, after many chat sessions. These include details such as financial goals, life events, employment history, etc. Now this user invests in a cryptocurrency and expresses excitement about diversifying more assets into crpyto. 
Using \textsc{SemanticCommit}, we add this piece of information to the list, and the system detects potential semantic conflicts which may require human review (Figure~\ref{fig:workflow}). A user clicks the ``Make Change'' button, which adds a new piece of information (deducing that it should do so, which is not always done), detects conflicts, then proposes changes to ensure the memory remains consistent with the new information. One line it proposes deleting entirely, another it rewrites, and others it flags for review. 

Notice how semantic conflict detection leveraged the LLM's general knowledge: a mention that the user likes Warren Buffet's investment strategies is highlighted as a potential conflict. Buffet, a famous investor, avoids cryptocurrencies and has  declared them ``rat poison squared'' \cite{kim2018buffett}. Clicking the Let AI Propose Change button on the \textit{local} information, a slight rewrite is proposed where the claim is softened (Step 4 in Fig.~\ref{fig:workflow}).

\begin{figure}
    \centering
    \includegraphics[width=\linewidth]{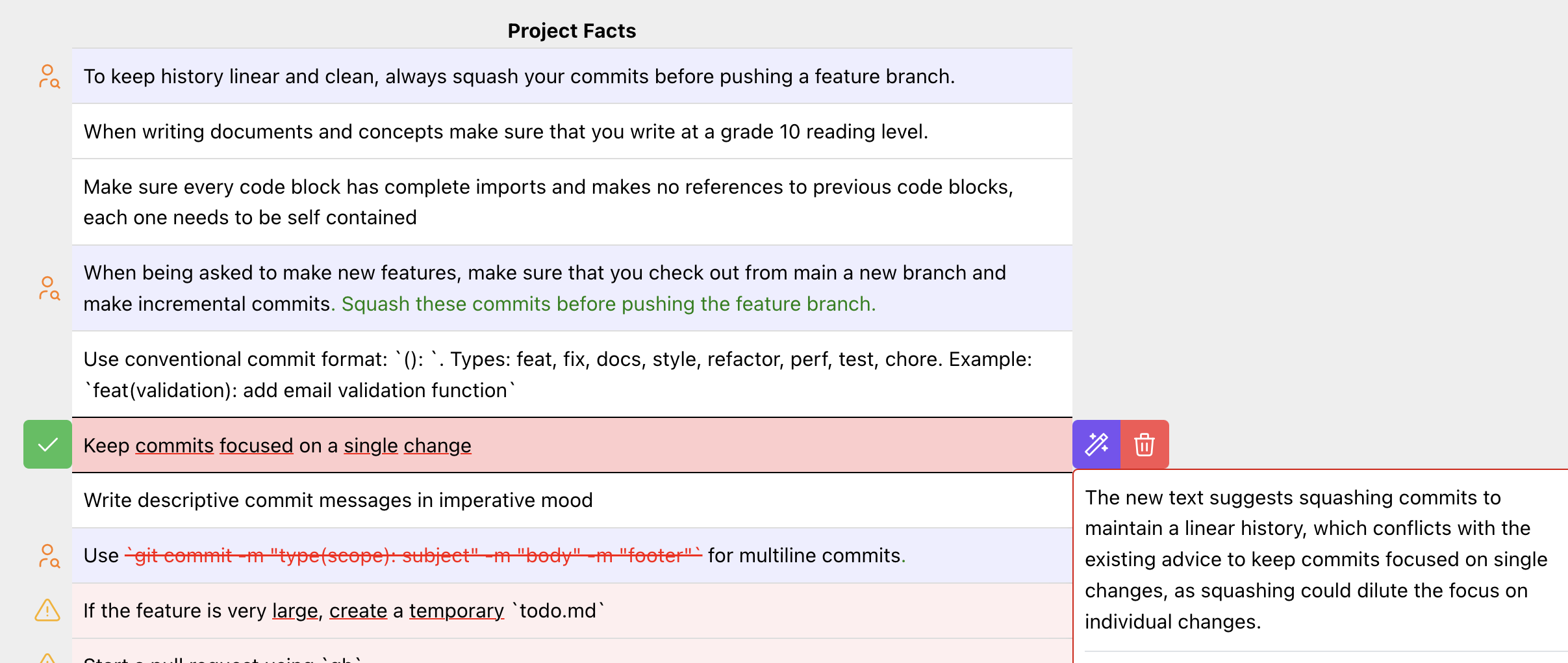}
    \caption{Cursor Rules~\cite{terragni2025future} adapted from the Instructor library~\cite{instructor_cursor_rules}, loaded into  our \textsc{SemanticCommit} UI. The user has added a new directive to squash commits before pushing a feature branch. The system adds the new rule to the top, makes a clarifying revision, and flags other lines as potential conflicts. One change is in error, which the user can quickly spot and revert.}
    \label{fig:cursor-rules}
\end{figure}

\subsubsection{\textbf{Updating Rules for an AI Software Engineer.}} Consider a user has a list of Cursor Rules, describing how an AI software engineering agent should behave in a code repository (Figure~\ref{fig:cursor-rules}). (Here we use real cursor rules adapted from \texttt{Instructor} API's open-source repository~\cite{instructor_cursor_rules}.) The user adds a new directive, common to software engineering practice: ``To keep history linear and clean, always squash your commits before pushing a feature branch.'' \textsc{SemanticCommit} highlights ``Keep commits focused on a single change'' in red, indicating direct conflict, and ``If the feature is very large, create a temporary \`{}todo.md\`{}'' in pink, indicating an ambiguity. The first is unclear how to resolve: removing it seems unwise, but keeping it unchanged incurs confusion. The AI has also added a mention of squashing commits, after the line, ``When being asked to make new features, make sure that you check out from main a new branch and make incremental commits.''

\subsubsection{\textbf{Changing a Game Design Document.}} \label{example-gdd} Finally, imagine a game designer has a design document for a game set on Mars, which an AI agent implements. After some playtesting, they decide that Mars is overused in sci-fi narratives, and communicate that they want to switch the setting to Venus. Here, the AI has estimated that the change is significant enough to request further clarification from the user before continuing:

\begin{center}
    \setlength{\fboxsep}{0pt} %
    \fbox{\includegraphics[width=0.6\linewidth]{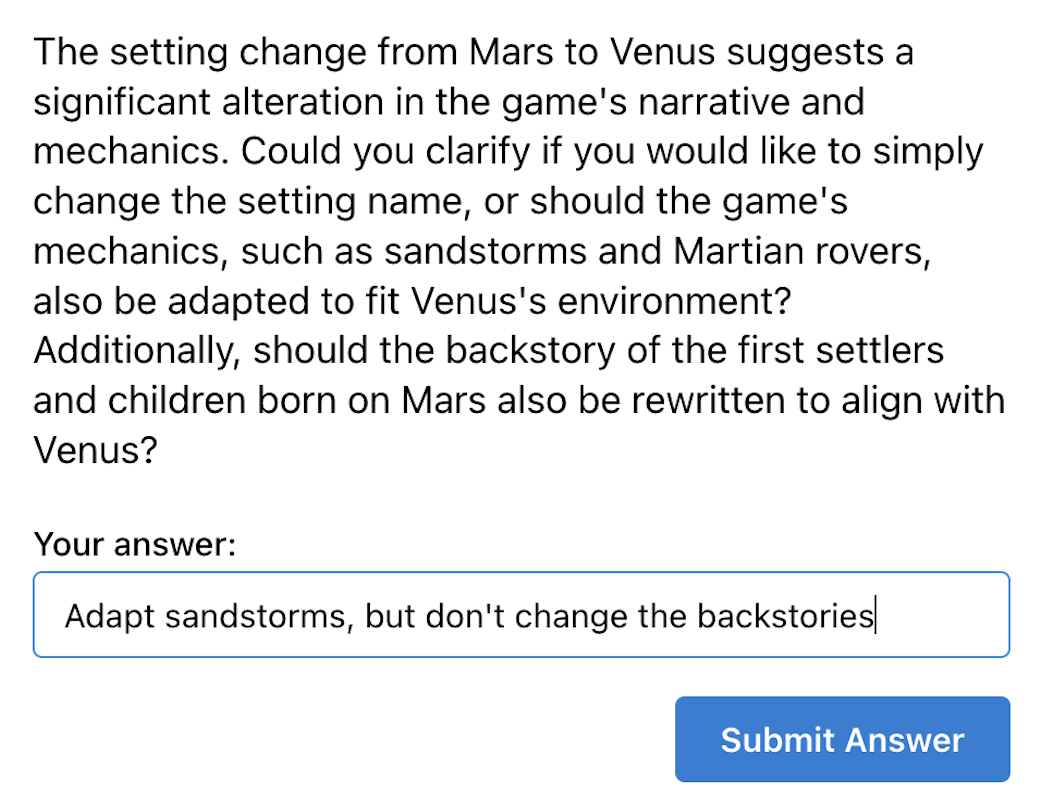}}
\end{center}

The user provides clarification, and conflict detection proceeds. The AI makes the most obvious changes---changing the term ``Mars'' to ``Venus,'' mainly---while flagging other potential conflicts for review. A subtle semantic conflict, that Mars has sandstorms but Venus does not, is detected and changed to a more generic ``storm'', steered by the user's clarification:

\begin{center}
    \includegraphics[width=\linewidth]{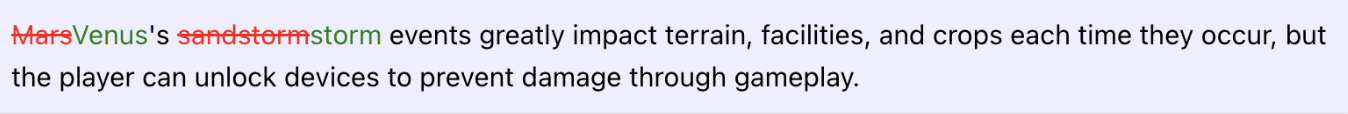}
\end{center}

\noindent These examples illustrate that conflicts: a) may require general world-knowledge to detect, b) may be hard to resolve, and c) \textit{how} to resolve a conflict can be a matter of creative decision-making. Resolving even a single change accurately is important, as unresolved conflicts can cascade as more changes are made. Using this system, we also \textit{learned} why some conflicts occur---the Buffett example above was not something we were aware of---or could be forced to reckon with second-order effects, such as re-thinking the sandstorm mechanic to better fit the  planetary conditions of Venus. 

Note finally that our system \textit{does} may mistakes---conflicting information can be missed, as our benchmarking shows; conflict detection and retrieval are stochastic; reasoning can sometimes be superfluous; and in practice, some knowledge base domains can benefit from adding a temporal feature to information (i.e., a limited duration where a rule holds). However, we believed the system was strong enough to run a user study in order to better understand where further efforts should be directed.

\subsection{Implementation}

The UI of \textsc{SemanticCommit} is implemented in React and TypeScript, with a Flask Python backend for our knowledge graph-based retrieval architecture (described in Section~\ref{tech-eval}). We iterated on prompts using ChainForge~\cite{arawjo2024chainforge} by setting up an evaluation pipeline against our benchmarks, which allowed us to observe the effects of prompt changes and model choices. There are \textit{many} prompt-based functions in \textsc{SemanticCommit}, from the user intent router, to conflict detection, local and global revision, underlining ``highly conflicting'' words, and suggesting resolution strategies. We chose GPT-4o for performance and latency reasons, as it performed optimally against our benchmarks. Further details on our development process and system are provided in Supplementary Material.

\section{Back-End for Semantic Conflict Detection}\label{tech-eval}

\begin{figure*}[t!]
	\centering
	\begin{subfigure}[t]{0.23\textwidth}
		\centering
		\includegraphics[scale=0.4]{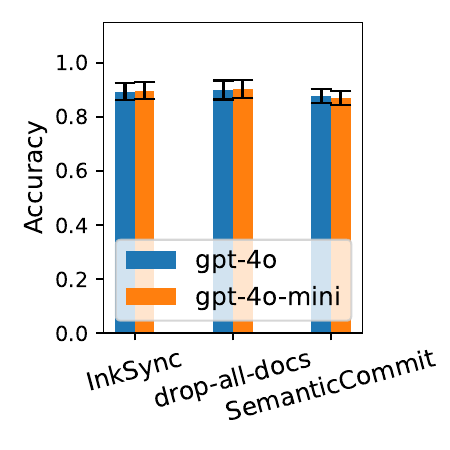}
		\vspace*{-0.5em}
		\caption{Accuracy (Mean ± StdDev)}
	\end{subfigure}%
	\hfill
	\begin{subfigure}[t]{0.23\textwidth}
		\centering
		\includegraphics[scale=0.4]{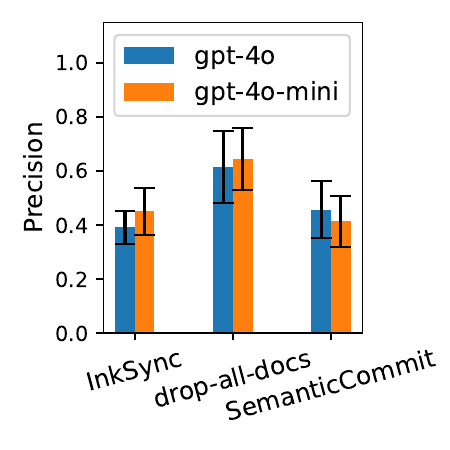}
		\vspace*{-0.5em}
		\caption{Precision (Mean ± StdDev)}
	\end{subfigure}
	\hfill
	\begin{subfigure}[t]{0.23\textwidth}
		\centering
		\includegraphics[scale=0.4]{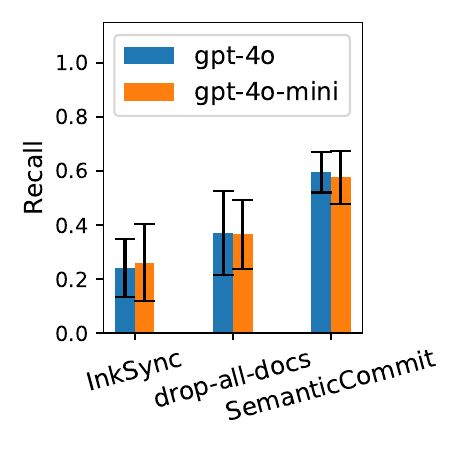}
		\vspace*{-0.5em}
		\caption{Recall (Mean ± StdDev)}
	\end{subfigure}%
	\hfill
	\begin{subfigure}[t]{0.23\textwidth}
		\centering
		\includegraphics[scale=0.4]{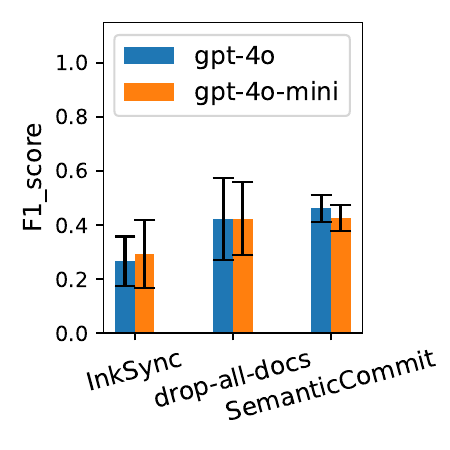}
		\vspace*{-0.5em}
		\caption{F1 Score (Mean ± StdDev)}
	\end{subfigure}
	\vspace*{-1em}
	\caption{Comparison of \experiment using a knowledge graph with PageRank relevance assessment and then classification to two baselines: (i) \textsc{DropAllDocs}: takes all documents in context to classify them without a retrieval stage; and (ii)  \textsc{InkSync}~\cite{Labin2024Beyond} implementation, reformulating the prompt to our context. The comparison is across all benchmarks in Table~\ref{tab:benchmark_stats}, averaged with st. dev. bars, for the \texttt{GPT-4o} and \texttt{GPT-4o-mini} models. Our method, \emph{kg-pagerank}, achieves higher recall with similar accuracy.}
    \label{sec:results-experiments}
    \vspace{-0.2em}
\end{figure*}

We implement a back-end system to drive the interface of \textsc{SemanticCommit}. 
The back-end's primary goal is to enable conflict resolution at scale. 
During early prototyping, we found that simple methods---giving the entire context to the LLM, or generating string-replace operations~\cite{Labin2024Beyond}---were prone to missing conflicts. These techniques rely on a single prediction, which takes as input the entire memory store and produces as output either a reformulated version or a set of suggested edits. Simple methods like rewriting the entire document frequently introduced superfluous changes, unrelated to conflict resolution, take a substantial input context, and can have large latency due to output sizes. %

To tackle the aforementioned limitations, we implement the back-end using a retrieval-augmented generation (RAG) approach~\cite{gutierrez2024hipporag} consisting of two phases: pre-processing and inference. 
The pre-processing phase constructs a \emph{knowledge graph} (KG) by extracting entities from a collection of input documents in the memory store and linking them. Each of the entities keeps track of the relevant document from which it was extracted. 
The inference phase detects %
semantic conflicts using a multi-stage information retrieval (IR) pipeline. 
The IR pipeline 
takes as input an edit action, whether it is an insertion or a modification to the memory store, and produces as output a list of chunks of information in conflict. 
It contains two stages: 
(i) \emph{retrieval}: finds relevant chunks of information using the KG in a single-step to avoid error propagation. 
In order to minimize relevance assessment issues, we apply a PageRank-based relevance ranking over the KG, akin to HippoRAG~\cite{gutierrez2024hipporag}; and
(ii) \emph{conflict classification}: identifies from the retrieved chunks of information which ones are in conflict with the edit. 

In the rest of this section, we give an overview of our design considerations and their rationale through an technical evaluation. We highlight that our prototype back-end system, achieves higher recall than the simple methods with similar accuracy. 

\subsection{Technical Evaluation}
\label{subsec:empirical-evaluation}

Our goal is to technically validate key aspects of our design decisions. 
We compare our end-to-end system against two simpler methods: (i) \textsc{DropAllDocs}, which adds all documents to the context for conflict classification; and (ii) \textsc{InkSync}~\cite{Labin2024Beyond} which generates a JSON list of string-replace operations. These comparisons allow us to analyze the impact of separating conflict detection from resolution, separating retrieval from conflict classification, and evaluating the performance of different LLMs.

\subsubsection{Benchmarks}\label{subsubsec:benchmarks} As part of our evaluation, we contribute four small benchmarks on three distinct domains:

\begin{itemize}
	\setlength{\itemsep}{2pt}
	\setlength{\parskip}{0pt}
	\setlength{\parsep}{0pt}
	\item \textbf{Game Design}: We use two game design documents. The first is from Labyrinth~\cite{fox1986labyrinth} by LucasArts (1986). The second includes excerpts from an original by one coauthor, describing a fictional game set on Mars about the first generation of children born there. The documents are chunked into paragraphs and referred to as the \emph{Labyrinth} and \emph{Mars} datasets, respectively. 
	\item \textbf{Financial Advice AI Agent Memory}: AI agent memory in the style of OpenAI's ChatGPT memories, about the investment strategies, financial situation, and background of a fictional college student living in South Korea (prepared by a South Korean coauthor based on their early 20s experiences). We refer to this as the \emph{FinMem} dataset.
	\item \textbf{Coding Assistant Rules}: Rules for the Cursor IDE~\cite{cursor_rules_for_ai}, which are intent specifications for coding assistants. A subset of the rules was adapted from the awesome-cursorrules GitHub repository. We refer to this as the \emph{CursorRules} dataset.
\end{itemize}

\begin{table}[t!]
	\centering
	\begin{tabular}{lrrl}
		\toprule
		\textbf{Benchmark} & \textbf{Ch} & \textbf{M} & \textbf{CS (Min, Median, Max)} \\
		\midrule
		\emph{Labyrinth}   & 35 & 17 & (0, 4, 10) \\
		\emph{Mars}        & 30 & 25 & (0, 2, 14) \\
		\emph{FinMem}      & 30 & 17 & (0, 4, 10) \\
		\emph{CursorRules} & 65 & 19 & (0, 3, 25) \\
		\bottomrule
	\end{tabular}
	\caption{Benchmark details including number of chunks (Ch), number of prepared modifications (M), and conflict statistics (CS) (min, median, max) across modifications.}
	\label{tab:benchmark_stats}
	\vspace*{-2.8em}
\end{table}

\noindent For each of these datasets, we introduce possible updates as insertions or modifications to chunks of information, all of which intentionally introduce varying yet targeted numbers of conflicts. 
Table~\ref{tab:benchmark_stats} summarizes each of the benchmarks including the number of chunks, the number of prepared modifications to apply as part of the benchmark, and the conflict statistics, i.e., min, max, and median, that these modifications lead to. 
These initial benchmarks served as a foundation for prototyping our approach and preparing user studies. We believe that developing similar, yet more sophisticated benchmarks is a valuable direction for future research.

\subsubsection{Models}

We evaluated the following LLMs by OpenAI: \texttt{GPT-4o}, \texttt{GPT-4o-mini}, and \texttt{3o-mini}, all with context windows large enough to accommodate each of the benchmarks.

\subsubsection{Experiments and Discussion} %

We compare our approach with the two baselines: 
\textsc{DropAllDocs} and \textsc{InkSync}. 
We run end-to-end on the four benchmarks using \texttt{GPT-4o} and \texttt{GPT-4o-mini} and report the mean ± stddev for
accuracy, precision, recall, and F1 scores for the three approaches in Figure~\ref{sec:results-experiments}. 

Our results show that \textsc{SemanticCommit} achieves higher recall ($1.6\times$ and $2.2\times$ higher) than \textsc{DropAllDocs} and \textsc{InkSync}, respectively, while retaining similar accuracy. This better addresses user preferences observed in our pilot studies and mentioned in related work, reducing risk of false negatives (Section~\ref{rw-nli}). 
Additionally, our system matches the F1 score of \textsc{DropAllDocs}, outperforming \textsc{InkSync} by $1.6\times$. While its \textit{precision} is comparable to that of \textsc{InkSync} and $1.6\times$ lower than \textsc{DropAllDocs}, we consider this an acceptable trade-off given our emphasis on maximizing recall. Note that our benchmarks are rather skewed with highly targeted conflicts (on average, only a few ground truth items in conflict when integrating new information), and accuracy can be misleading in such a setup, as assigning non-conflict to all documents would still yield high accuracy. %

Overall, \textsc{InkSync} %
performs worst likely due to its combination of both conflict detection and resolution in a single prediction. 
In contrast, both \textsc{SemanticCommit} and \textsc{DropAllDocs} benefit from task decomposition, achieving similar F1 scores. %
\textsc{SemanticCommit}'s additional decomposition intro retrieval and conflict classification enables independent optimization contributing to the higher recall. 
This decomposition proves beneficial even when it is possible to fit all documents into the context window, as we observe worse conflict classification as the false positive rate (FPR) increases. Filtering down the chunks of information remains preferable. 

We selected \texttt{GPT-4o} for its slightly better performance, comparable latency to \texttt{GPT-4o-mini} and for being twice the speed of \texttt{o3-mini}. 
Additional details on FPR sensitivity and a comparison with \texttt{o3-mini} are provided in our Supplementary Material.

\section{User study}
\label{user-study}

To understand how users integrate new information in practice, we conducted a controlled within-subjects study with mixed methods, comparing \experiment with a baseline interface. We had the following research questions:

\begin{itemize}
    \item Which interface affordances do users prefer (use most often) when performing an integration of new information?
    \item How do users think through the process of integrating new information into an AI's existing memory store, with regards to detecting and resolving potential conflicts? 
    \item Does \textsc{SemanticCommit} make users feel more in control of the integration process, over a more open-ended one?
    \item Does \textsc{SemanticCommit}'s required manual review increase user workload compared to a more automated method?
\end{itemize}

We felt it worthwhile to have a baseline interface to better understand: 
1) any interface affordances our structured environment might miss, compared to an open-ended one; 
2) how users might currently use popular AI-based tools to handle the process of integration, in the absence of targeted support. 
We chose OpenAI's ChatGPT UI as a baseline for three reasons: 
(i) it is likely familiar to users. 
(ii) it provides a ``Canvas'' view for document editing assistance, where users can select text and ask GPT to rewrite it, or chat with an AI to make global edits; and 
(iii) similar interfaces like Anthropic Artifacts tended to rewrite the %
specification entirely, and did not offer Canvas's ``diff'' view to allow for a fair comparison.\footnote{We focused on AI-assisted conditions because our ultimate goal (and anticipation) is that AI will keep track of user intent, especially as the intent specification grows lengthy and unwieldy. Even within our toy benchmarks, we encountered how time-consuming conflict detection can be: manually identifying conflicts for a single new piece of information could easily take 10 minutes, if one was being precise.}

\minipar{Participants.} We recruited 12 participants (7 female, 5 male) through the mailing lists of two research universities and one multi-national technology company. 
All the participants were familiar with GenAI tools. Ten participants used GenAI tools daily, and the other two at least weekly. 
ChatGPT was the most commonly used tool, alongside others, e.g., Gemini, MS Copilot, Claude, Perplexity, and Deepseek. Seven participants had previously used Canvas-like tools, and eight had used persisting memories (or preferences) with AI tools. Of these eight, four participants actively manage their memories either by adding, editing, or deleting them. Participants received a \$25 Amazon Gift Card as compensation.

\minipar{Tasks.} We adapted two intent specifications from our benchmarks. 
We chose the \textit{Mars Game Design Document} and \textit{Financial Advice AI Agent Memory},
as these tasks mapped to the two paradigmatic types of intent specifications covered in Sections~\ref{motivation} and \ref{related-work}: design documents, and AI memory of the user. 
We ensured each list was 30 items long as our pilot studies suggested this was long enough that manual detection starts to become unwieldy (users need to scroll up and down the document), but short enough that participants could become familiar in a short period. 
For each task, participants were tasked with integrating three new pieces of information into the memory, one at a time (``sub-tasks''). %
We told participants to only change pieces of information that conflict with the new information, and that otherwise they were free to make additions, edits, and deletions as they saw fit.
One of our tasks directly asks users to imagine they are an information management system that is managing memories about the user, in order to mimic how automated memory management systems will need to be conservative in changing information. 
More details on our tasks are provided in Supplementary Material. %

\minipar{Procedure.} To enable easy access to the \experiment, we hosted the tool online, allowing participants to access it via their web browser. For access to \control, we provided credentials for a ChatGPT account specifically created for the study to control for model and feature differences. With participant consent, we recorded audio and screen-casts, and participants were encouraged to think aloud during the study. In each study session, the participant completed one of the two tasks each (each task containing 3 sub-tasks) using both the tools. Both the order of task assignment and the order of tool assignment were counterbalanced and randomly assigned. Before each task, participants received a tutorial on the assigned tool and were given five minutes to explore it using a test document. We also provided a summary of the task document and time to read through it before starting. Each condition had a time limit of 15 minutes, after which the participant completed a post-task survey. After both tasks were completed, participants filled out a final survey to compare the two conditions. We then conducted an informal interview to better understand each participant’s experience.

\minipar{Measurement and Analysis.} For each task, we measured the success or failure of each sub-task the participant was required to perform. A sub-task was considered a failure if the participant was unable to complete it within the time limit. For condition using \experiment, we recorded all instances of \emph{edits}, \emph{check for conflicts}, \emph{make change}, \emph{local}, and \emph{global resolution} actions using telemetry. In the post-task survey completed after each task, we collected self-reported NASA Task Load Index (TLX) scores, Likert-scale ratings for ease of use, and responses on how well the AI helped participants identify, understand, and resolve semantic conflicts. In the post-study survey, completed after both tasks, we recorded participants’ self-reported tool preferences and a modified NASA TLX focused on comparing their experiences between the two tools. For qualitative analysis, the first author performed open coding on participant responses and audio transcripts to identify themes, which were used to interpret the qualitative results. To measure statistical significance, we used Mann–Whitney–Wilcoxon tests and report the p-values.

\subsection{Findings}

\begin{figure}[t!]
    \centering
    \includegraphics[width=0.9\linewidth]{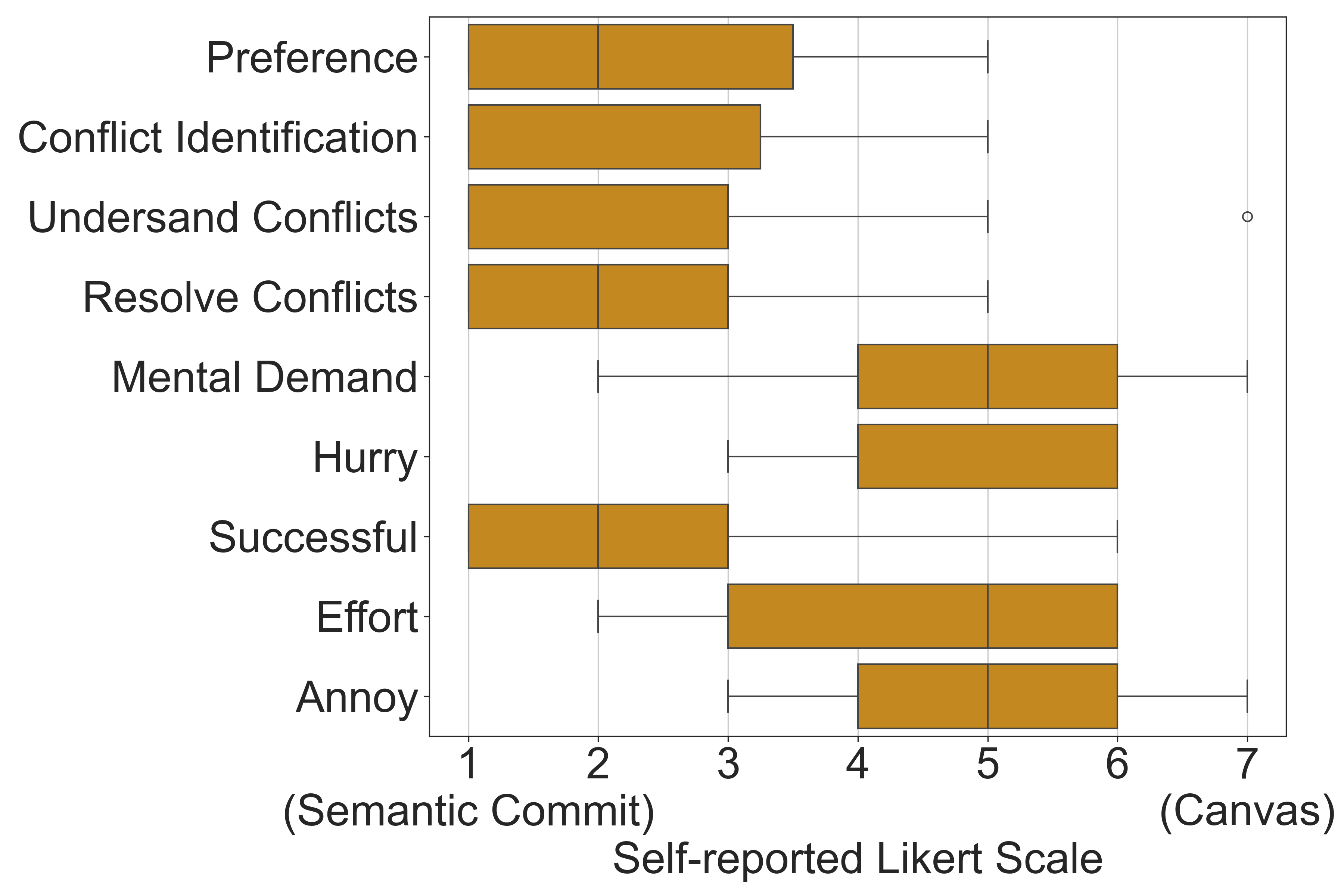}
    \vspace{-1.0em}
    \caption{Participants' self-reported cognitive load and preference scores that directly compare the two conditions.}
    \vspace{-1em}
    \label{fig:sc-post-study}
\end{figure}

\boldsec{Preferred Workflow}

Participants employed distinct workflows with each tool. We recount three characteristic workflows of \experiment first, then compare to user behavior in  \control. 

\minipar{Impact analysis first.} Six participants (P1, P2, P4, P5, P9, P10) always began with \textit{Check for Conflicts}, gaining insight on the impact of the change before integrating any changes. Participant P2 explained why they prefer \emph{check for conflicts} by saying \sayit{I really like the \emph{check for conflicts} action -- it still gives me control, and it feels collaborative instead of me kind of scrolling through the whole thing and trying to find it [referring to \control]. It highlights points of issues where I can plug this in.}
Participant P7 explained, \sayit{I know where to make the edit, but I will use the global check so that I can find other places I might have to change}.
All but one participant in this group proceeded exclusively with localized edits afterward. %

\minipar{Immediate changes with conflict review.} Five participants (P3, P6, P7, P11, P12) always started the task with the \emph{Make Change} feature to see the conflicts and the potential changes at once. They then followed up with local changes. P3 said \sayit{This one has a lot of changes, so I’m going to use the global option. I’m just going to \emph{make change}, and then figure out what to keep.}

\minipar{Skim to resolve false positives before proceeding.} A method adopted within the two workflows, four participants (P3, P6, P9, P10) using \experiment first quickly perused all the conflicts to resolve the false positives\footnote{Participants considered ``false positives'' as the conflicts flagged by the system that, in their judgment, did not require meaningful intervention. This highlights the subjective nature of conflict interpretation.} %
and then proceeded to spend time resolving the \emph{actual} conflicts. %

\minipar{In \control, users instead lean heavily on global rewrites.} When using \control, eight participants (P1, P2, P4- P6, P10 - P12) predominantly utilized global prompts, instructing ChatGPT to perform edits throughout the entire document, while four participants preferred starting with global edits and subsequently performing local rewrites by selecting specific lines. As we recount below, this behavior intersected with frustrations from lack of control and the metacognitive demands~\cite{tankelevitch2024metacognitive} of prompting. 

\minipar{Workflow choice can depend on context.} When asking participants how they pick between local vs. global resolution, they gave two major reasons---complexity of change and familiarity with the document. For example, P9 mentioned they would use global resolution techniques when they perceive the impact is higher---\sayit{There is a lot of information here, it is much harder to go through it one by one. So I wanted to check for all the conflicts with the doc and then change it [collectively].} The choice also depends on how familiar they are with the contents of the document. P12 said \sayit{And I’m gonna go to [{\experiment}] and put this as a global change. And I’m gonna say, first check for conflicts before making a change because I haven’t read the complete document thoroughly.}

\boldsec{Improved ability to catch semantic conflicts}

Nine participants (P1 - P4, P7, P8, P10 - P12) explicitly stated that \experiment was better at identifying conflicts compared to \control. In the post-study survey ranking, participants additionally report a higher level of \emph{task success} with \experiment compared to \control ($\mu{=}2.42; \sigma{=}1.5$, where $1$ indicates full preference for \experiment), higher levels of success in \emph{identifying semantic conflicts} ($\mu{=}2.08; \sigma{=}1.5$) and in \emph{understanding semantic conflicts} ($\mu{=}2.25; \sigma{=}1.95$). As P4 noted \sayit{It feels like you can identify inconsistencies easier in [{\experiment}], which is what I liked a lot. So I favor that more. I'd feel I'd be a lot faster at getting work done.} %

This preference stemmed from two primary reasons. First, six participants (P2, P3, P4, P8, P11, P12) explicitly mentioned that when using \experiment, the granularity of information and the red-colored highlights enabled easy conflict identification. P12 explained this in terms of context for the AI by saying \sayit{I think the [{\experiment}] tool is great in finding conflict, that's because it discretizes information, it’s much more granular. It doesn’t club all the context together.}  Second, except P2, all the other participants heavily relied on the \textit{rationale} provided by \experiment when understand why a conflict occurred. P8 explained this by saying \sayit{With [{\experiment}]... %
there is stronger explanation provided as to why that conflict is occurring.}

\minipar{Inconsistent conflict detection in \control leads to frustration and flailing.} In contrast, nine participants (P1 - P3, P5 - P9, P11) noted that \control often missed conflicts or failed to understand the changes they wanted to make. Across 18 cases involving 10 participants (P1, P2, P3–P7, P9–P12), \textit{\control failed to detect even a single conflict during the task.} In 9 of these cases, participants accepted the results without further checks; in the others, they had to either manually spot the issues or retry with more specific prompts. We highlight some of the observations below.

In one instance, P5 had explicitly asked \control to find conflicts in the document. When the tool failed, the participant manually pointed out a conflict by quoting the text, and the AI model came up with a convoluted reason as to why it was not a conflict. P5 retorted by saying \sayit{It is giving me an excuse.} %
In a different task, P5 exclaimed \sayit{Looks like it just added one statement, and there is no conflict. [5 seconds later] Oh wait! the GameBoy aesthetics is conflicting}---catching a false negative manually in real time. %
In another instance, P9 prompted the \control tool three times to identify conflicts and make a change, but each attempt failed. Frustrated, they exclaimed, \sayit{It didn’t change it the way that I wanted. Maybe I’ll delete this and do it myself and specify what I want to be changed} before proceeding to manually make the change.

There were eight instances with six participants (P1, P2, P5, P7, P11, P12), where \control drastically changed the contents of the document either by replacing all the contents or by making heavy modifications. We then instructed the participants to restore to a previous version using version history.

\begin{figure}[t]
    \centering
    \includegraphics[width=0.7\linewidth]{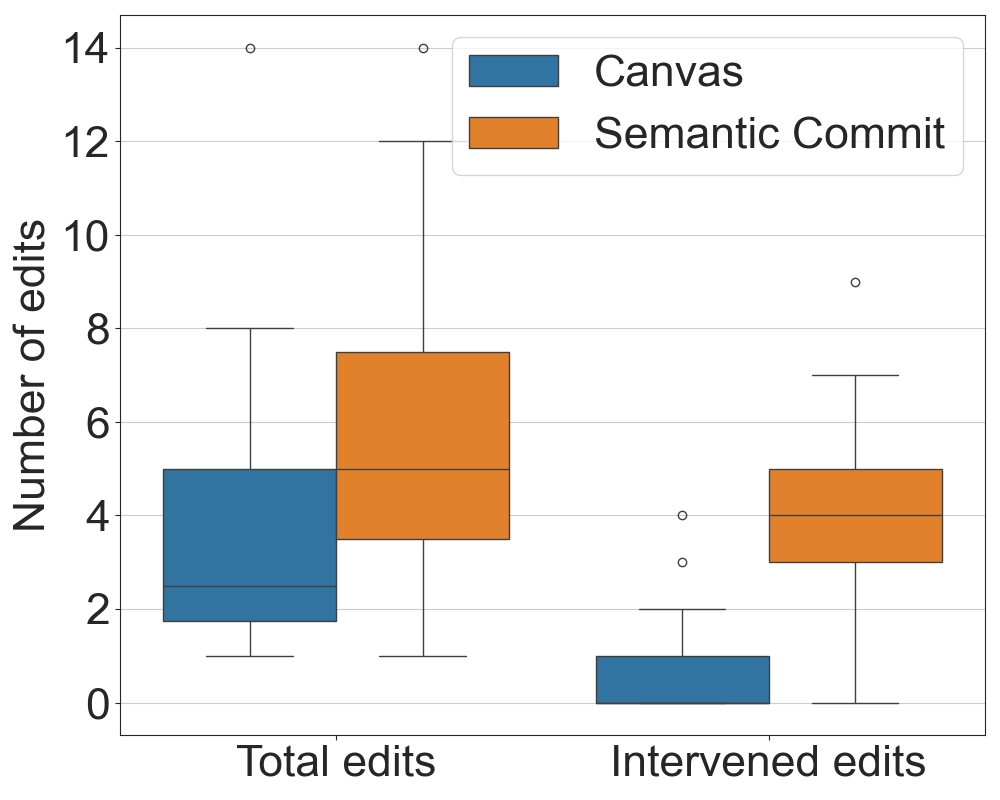}
    \caption{Participants using \experiment made significantly more edits and intervened edits compared to \control.}
    \label{fig:sc-edits}
    \vspace{-1em}
\end{figure}

\boldsec{Greater sense of control with \experiment}

A recurring theme among participants was the strong sense of control they felt while using \experiment. Nine participants (P2, P3, P4, P6, P7, P8, P10, P11, P12) explicitly mentioned that \experiment offered them more control over the integration process compared to \control. In the post-study survey ranking, participants additionally report a higher level of \emph{control} with \experiment compared to \control ($\mu{=}2.08; \sigma{=}1.36$, where $1$ indicates full preference for \experiment), as well as a higher level of success in \emph{resolving semantic conflicts} ($\mu{=}2.17; \sigma{=}1.34$). This perception of control emerged due to several reasons mentioned below.

\minipar{Granular insights into conflicts}: Six participants (P2, P3, P4, P8, P11, P12) emphasized that the fine-grained presentation of information in \experiment made it easier to identify, understand, and resolve conflicts---particularly for localized edits. The piece-by-piece %
breakdown gave users a clear sense of what was being altered and why. As P11 explained, \sayit{you have some concept of a line—every element is aligned, so you probably have more granularity to control the elements that are being changed. That was really nice... %
I never had to worry that the entire document is going to get changed here and there.} 
This precision allowed participants to maintain a stronger grasp over %
editing %
and focus their attention where it mattered. %

\minipar{Conflict reasoning encourages critical reflection}: The tool’s detailed breakdown of conflicts and its reasoning behind proposed changes encouraged users to think more critically about edits. P12 described how this led them to reevaluate parts of the content they might have otherwise overlooked: \sayit{So yeah, [{\experiment}] improved the conflict finding even more... %
there were some parts in the document I would have ignored if I was doing it on my own. I wouldn’t have considered some graphic design aspects of the game, but [{\experiment}] provided its reason on why it has raised this as a conflict made me reconsider my decision. I like that part, because I would have easily ignored it, and that would have led to more iterations with more discussions.}

\minipar{Forced review enhanced sense of control over process}: Another factor that reinforced a sense of control was the editing workflow itself. Unlike \control, which applied changes automatically, \experiment required users to first review conflicts, make changes, and then manually click the resolve button to validate them. %
This structure helped participants feel like they were directing the process. As P10 observed, \sayit{In [{\experiment}] it was a step by step process to see the conflict, before making any changes whereas in [{\control}] there was no decision making on my behalf and it did the changes all by itself whether I agree with it or not.} %
Similarly, P4 noted, \sayit{Making changes [with {\experiment}] was my favorite, because it walks you through every line, highlighting recommendations like revise, delete, change, add, or nothing.} 

This workflow—of reviewing conflicts, followed by local and/or global resolution—also could make the task feel more collaborative. Three participants (P1, P2, P5) described the process with \experiment as collaborating with AI. P2 said \sayit{With [{\experiment}] you could ask it to look for conflicts. So you're sort of partnering like it would get the conflicts for you, and then you would move through them systematically... %
I felt like with [{\control}], you didn't have that middle ground. It was either make the change or don’t.}

\minipar{Loss of control breeds insecurity.} Due to \control not identifying conflicts and understanding instructions from participants, combined with sudden and drastic changes to the document, eight participants (P1 - P6, P10, P11) explicitly mentioned they have doubts and insecurity when using \control to make any changes to a document. P2 said \sayit{Using [{\control}] was really uncertain. You know, you just kind of felt like you're guessing, and you didn't know what was gonna happen.}  P6 also explained this by saying \sayit{The downside of [{\control}] will be you just take it as it is, so you may not notice there’s a part that should or shouldn’t be changed. You may just skip it, pass it, and never notice the mistake the AI tool made.}

\minipar{Responsive UI with many local resolution options}: Participants also appreciated the responsive nature of the interface during local resolution. As P11 described, \sayit{The [{\experiment}] tool I found quite intuitive, especially with the responsive nature where you put your mouse on it and there’s a color code, and there’s a green resolve button. The right-hand side gives you options to revise, reject, delete, edit, or suggest a new revision, etc. That is really good.} %

\minipar{Ease of reversibility}: Like diff interfaces, participants also valued the ability to manually review changes and locally undo or dismiss them. P11 noted the friction in \control’s reversal process: \sayit{With [{\control}], if you want to reject changes, then you probably have to undo and restore to the previous version, which seems a little cumbersome. It’s not as simple as in [{\experiment}] where you could accept a change or reject right there in that line.} 

\minipar{Tradeoffs between control and efficiency}: While many appreciated the explicit approval mechanisms in \experiment, a few also noted potential tradeoffs. P3 acknowledged that the confirmation steps could feel excessive in low-conflict scenarios: \sayit{I think sometimes it was overkill, if there were a pretty low number of conflicts detected. But otherwise, I think it was nice to confirm.} P12 framed this as a tension between control and usability: \sayit{I think it’s important to do if you want finer control, but it really depends on the application you want to package it as. If you want better user experience, and you do not want them to spend more time, you would have to give them less control.}

\boldsec{Perceived cognitive load}

In the post-study survey, participants’ preferences were measured using a 7-point Likert scale, where 1 indicated a strong preference for \experiment and 7 indicated a strong preference for \control. Participants reported slightly higher levels of \emph{mental demand} ($\mu{=}4.67; \sigma{=}1.56$), \emph{hurry} ($\mu{=}4.75; \sigma{=}1.14$), and perceived \emph{effort} ($\mu{=}4.5; \sigma{=}1.62$) when using \control compared to \experiment. They also reported slightly greater feelings of \emph{annoyance} ($\mu{=}5; \sigma{=}1.2$) with \control.

However, when comparing post-task questionnaires, we observed no statistically significant difference between conditions regarding mental demand, sense of hurriedness or frustration, effort exerted, or perception of success (all p-values are 0.45 and above). This null result was surprising to us, as we had expected \textit{higher} workload in the \textsc{SemanticCommit} condition due to the increased demand as users manually click to resolve conflicts. 

\boldsec{Task time and completion rates}

On average, participants took \emph{}{4 minutes and 7 seconds} (\emph{$\sigma{=}$ 117 seconds}) to complete tasks using the control tool compared to \emph{5 minutes and 41 seconds} (\emph{$\sigma{=}$ 123 seconds}) using the experimental tool. This difference is statistically significant ($p{\approx}0.004$). It is important to note that task completion time does not capture task performance due to their exploratory nature, which encouraged participants to spend additional time holistically integrating document changes. We observed no significant difference in task completion rates between the two conditions. Four participants failed to complete one sub-task with \control compared to five participants with \experiment, with all failures attributed to insufficient time.

\boldsec{Participants made significantly more edits when using \experiment}

Measuring participant engagement in controlled lab studies is challenging. Counting \emph{edits}---with more edits typically indicating higher engagement---is useful, but  AI tools can easily automate extensive editing, reducing  reliability of metrics. To address this, in addition to studying number of edits overall (human- or AI-made), we also studied \emph{intervened edits}---edits explicitly triggered by participants one at a time, whether manual or with AI. These metrics give a more comprehensive picture. %

Participants using \experiment demonstrated significantly higher engagement across both measures. They made an average of 5.83 edits ($\sigma{=}3.21$), compared to 3.5 edits with control ($\sigma{=}2.85$; $p{\approx}0.001$). 
This contrast was even stronger for \emph{intervened edits}, where participants using \experiment averaged 4 edits ($\sigma{=}1.94$) per task, while participants using \control averaged just 0.65 ($\sigma=1$; $p{<}0.001$;~\autoref{fig:sc-edits}). 
Finally, when using \experiment, participants made an average of 2.93 \textit{localized edits} per task, significantly ($p{<}0.001$) higher than an average of 0.28 localized edits per task when using \control. 
Participants extensively used the different kinds of local resolution strategies such as \emph{revise}, \emph{add}, and \emph{delete} suggested by \experiment. 
These differences highlight the participants' willingness to make more edits when using \experiment. 
This also helps explain the higher average task completion time presented earlier---showing participants invested more time in understanding and making more deliberate changes. %

\boldsec{Participant trust and over-reliance}
Trust emerged as a complex and sometimes contradictory theme in how participants interacted with the AI tools. While many participants expressed skepticism toward AI-generated changes, their actual behavior revealed moments of over-reliance---particularly when changes appeared seamless or were not flagged as conflicts by the tool.

A majority of participants (P3, P4, P6, P7, P8, P10, P11, P12) explicitly stated that they did not trust the AI to make changes without their manual verification. As P10 firmly noted, \sayit{No, I don’t trust any AI blindly to make full and final changes to the result accurately. I always verify manually to spot any mistakes or misinterpretations by AI.} This sentiment reflects a baseline level of caution we expected the participants to carry throughout the tasks. When comparing the two tools, %
six participants (P1, P2, P5, P6, P11, P12) explicitly reported greater trust in \experiment over \control. They cited better contextual understanding and more transparency in the editing process as reasons for this preference. For example, P2 said, \sayit{With [{\control}] I was very skeptical. I don’t think I would trust it without doing a full read myself. With [{\experiment}], I trusted it more. I felt like it seemed to understand the context better. But no matter the tool, I need to make sure that everything was good, so I would still read it over again.}

Despite these widespread claims of skepticism, however, participants occasionally over-relied on both tools. As noted earlier, in nine instances where \control failed to identify any conflicts, participants accepted the output without further review. A similar pattern emerged with \experiment: five participants skipped reviewing parts of the document that were not flagged as conflicting. This points to a potentially risky dependency on the AI and underscores our decision to improve recall at the expense of precision---if the model fails to detect a conflict (false negatives), users may miss critical issues simply because they trust the system's silence.

\section{Discussion and Design Implications}

Based on our user study findings, we present design implications, discuss future work, and connect to relevant literature.

\subsection{AI agent interfaces should help users perform \textit{impact analysis}}

Our findings contribute to growing line of HCI research that emphasizes proactivity, presence, and just-in-time steering in AI agents acting on user's behalf \cite{kazemitabaar2025exploring, kazemitabaar2024improving, chen2024need, pu2025assistance, min2025feedforward, satyanarayan2024intelligence, shao2024collaborative}. The most surprising finding was participants' preference for performing \textit{impact analysis}: finding conflicts first before making any edits. Instead of automatically applying changes and prompting users to verify afterward (like \control), this suggests AI agent systems should encourage users to first understand the impact of the change and only \textit{then} choose to explicitly suggest and/or trigger changes. Our findings indicate higher trust and satisfaction when users actively initiate changes,  reducing uncertainty and increasing perceived control. %
Surprisingly, \textbf{the benefits from increased control seem to offset the cost of AI output validation}, as our results on perceived workload suggest. Not all users will use impact analysis in every context, but highlighting what aspects of an artifact will be considered and/or modified can help enhance user trust and control, especially in high-stakes situations.

This bears important implications for current AI agent interfaces, which tend to first let the AI make changes, and then have users \textit{validate} them. For instance, in AI-powered programming IDEs like Cursor and Visual Studio, the agent makes changes across documents and then presents the revisions for human review. Instead, \textbf{our findings call upon designers of AI agent systems to provide affordances for \textit{impact analysis}: helping users \textit{foresee} the impact or location of AI changes, \textit{before necessarily suggesting concrete changes}}. This reflects the principle of \textit{feedforward}~\cite{min2025feedforward, vermeulen2013feedforward} in communication theory---``a needed prescription or plan for a feedback, to which the actual feedback may or may not confirm'' \cite{richards1968secret}---where a communicator provides ``the \textit{context} of what one was planning to talk about,'' prior to talking about it \cite[p. 179-80]{logan2015feedforward}, in order to ``\textit{pre-test} the impact of [its output]'' on the listener \cite[p. 65]{griffin2006first}. This returns control to the user and explicitly separates \textit{retrieval} and \textit{generation}, two sides that are currently conflated in many agent interfaces. Such an affordance might also address a growing pain-point for users of SWE agents, where unrelated files and code are deleted without approval.\footnote{There are many examples of this, from  forums (\url{https://news.ycombinator.com/item?id=43298275}) to memes (\url{https://x.com/daniel_nguyenx/status/1909184057755496571}).}  

Note that impact analysis is not simply about pausing before enacting a change. It's also about weighing how extensive a change might be, the work required, and unintended side-effects. Users can use impact analysis to \textit{back out} of an in-progress change, before the damage is done or they are overloaded by AI slop---an \textit{AI resilient}~\cite{glassman2024ai, gu2024gptsm} affordance that helps users preemptively judge and respond to AI decisions. The reflective nature of impact analysis could also help users better understand potential conflicts, even inspire new ideas and areas for improvement. %

\subsection{Let the user walk the spectrum of control} 

When designing mixed-initiative systems~\cite{horvitz1999principles} where the users and AI collaborate, there is a trade-off between control (retaining it due to distrust in AI) and efficiency (completely delegating). %
\textsc{SemanticCommit}'s affordances for adjustable autonomy~\cite{bradshaw2003dimensions}, or blended agency~\cite{satyanarayan2024intelligence}, enabled the user to dynamically select their preferred balance between automation and manual oversight depending on the context, complexity of tasks, trust in the AI, or familiarity with the content, whereas users experienced loss of agency in the baseline condition. This suggests that AI agent interfaces should offer both highly controlled (step-by-step approvals like local resolution in \experiment) and streamlined (global changes) workflows to adapt to varying user needs. Our participants appreciated detailed explanations about identified conflicts and recommended resolutions, which empowered them to make informed decisions. Transparency also appeared to reduce anxiety and frustration, promoting critical evaluation rather than passive acceptance.

\subsection{Start global, then accelerate local review} 

We implemented a range of elements into \experiment{}, not knowing what users would prefer. We found that though users started globally, they preferred to then make local edits, and liberally used a range of local options---local steering, AI rewrites, etc---rather than global steering prompts and global resolution strategies. In the baseline \control condition, it was the exact opposite: users appeared resigned to global steering in chat and became frustrated by lack of granular control. This suggests \textbf{future interfaces for semantic conflict resolution should better support and accelerate local review,} rather than focusing on features for global steering after the initial interaction. The workflow of 4 participants to first dismiss false positives, and only \textit{then} focus on handling conflicts, was also telling. Interfaces might explore explicitly separating stages of ``double-checking'' AI outputs versus resolving.

\subsection{Future Work and Connections}

\boldsec{Interfaces and APIs for management of AI memory of user intent} %
We mentioned earlier that our intention is for \textsc{SemanticCommit} to become an API that helps users make ``semantic commits'': committing ideas and details to projects like we commit code, where the integration work is assisted by AI. Our UI was mainly a vehicle to see what users would do, were they given full control over the integration process. Left to their own devices to prompt chat models, our findings show that users are prone to miss conflicts or accept unwarranted rewrites of entire memory stores. Developers who utilize these simple one-shot prompting methods will be prone to similar failure modes. Tools like Claude Code provide users quick command-line directives to update memory, but simply append the directive to the end of the intent specification~\cite{claude_code_overview}.

What would a more \textit{assistive} command-line interface for memory updates look like? Could we automatically surface the conflicts that users care about, anticipating and correcting misalignments before they happen---potentially saving thousands of wasted inference calls? As AI agent systems grow in popularity, it becomes critical to explore interfaces and APIs that help users and developers alike \textit{manage}, \textit{inspect}, and \textit{update} AI memory of user intent in a manner that is \textit{non-destructive}, \textit{transparent}, and \textit{controllable}. 

The hard question is what to do when we \textit{don't} have the luxury of a graphical UI---when intent integration is an API, part of a larger system. When and how to raise conflicts for user review?
What rises to the level of ``direct conflict'' that must be addressed, versus an ambiguity that the AI could still proceed under? %
This goes back to our initial discussion on NLI and ambiguity, where human annotators had subjective differences in resolving conflicts \cite{chen2025onreference, jiang2022investigating}---in many cases, these differences emerged from humans holding \textit{different frames of reference}. To align conflict detection to specific users, we might consider two mechanisms---first, grounding acts like request for clarification \cite{shaikh2025navigatingriftshumanllmgrounding, shaikh2024grounding}, triggered contextually. Vaithilingham et al.~\cite{vaithilingam2024imagining} suggest that the benefits of negotiation increase with the level of abstraction: AI agents should engage users in  discussion for high-impact decisions, while avoiding it for low-impact ones. %
Second, a more passive mechanism might use memories to help model a particular user's classification of ``conflict,'' aligning it over repeated interactions \cite{shankar2024validates, shaikh2025aligning}. Future research could investigate how to align conflict detectors to \textit{specific} humans' ontological understanding of conflicts in their task domain.

\boldsec{Cognitive forcing functions to mitigate over-reliance} A line of research argues that to mitigate the risk of users becoming complacent or overly reliant on AI, systems should incorporate \textit{cognitive forcing functions}~\cite{buccinca2021trust, de2025cognitive} ---interface mechanisms that deliberately encourage active user involvement. In \experiment, we do this by requiring explicit user approval when a conflict is detected or a change is made by the AI. Such mechanisms foster sustained cognitive engagement and reduce the likelihood of critical oversights resulting from blind trust in AI-generated outputs.

However, mitigation of over-reliance is not elimination. %
Our work reflects the tension between automation and agency~\cite{heer2019agency, satyanarayan2024intelligence}, embodied by our efforts to enhance recall to reduce false negatives. %
Drawing user attention to conflicts---even ``ambiguous conflicts''---shows that users are liable to over-rely upon the AI to the extent of not checking any non-marked information. One further mitigation may be to mirror the kinds of divergences human annotators face when detecting conflicts~\cite{chen2025onreference, jiang2022investigating} by querying multiple LLMs in parallel and adopting a majority voting or ensembling scheme~\cite{tseng2024two}. The ``degree'' of conflict might then correlate with consistency and number of votes, and divergences in LLM judges could be visualized.

\boldsec{Interfaces to support requirements-oriented prompting}

Ma et al.~\cite{ma2024rope} introduced a process for prompting AI agents that focuses on supporting users in creating a good initial set of requirements. They argue that in the age of ``requirements-oriented'' prompting, HCI will need to focus on training users to be good requirements engineers. Although not entirely focused on requirements lists, our interface can help users update requirements to reduce conflicts, inconsistencies, and ambiguities. Future studies might explicitly study the performance of an AI agent following the user's intentions after changes are made. %

\boldsec{Semantic commits for long-form writing}

One of the impetuses for this work was inspired by the challenges a coauthor faced when performing developmental editing for a long fiction novel. Developmental editing \cite{norton2009developmental} %
assesses the overall content and structure of a document with regards to consistency, plot, and flow. Changed or removed scenes, even one-off conversations, could have impacts much later in a novel, and an author must keep all of this information in their head or manually reread to detect inconsistencies. 
A review by Zhao et al.~\cite{zhao2025making} found that little HCI research focused on helping writers perform developmental editing. In the future, NLI-like AI-powered interfaces might help writers of long documents detect and resolve inconsistencies that emerge as a result of revisions. 
Much like \textit{Portrayal}~\cite{hoque2023portrayal} shows writers birds-eye views of characters across a novel, might a similar interface help users to visualize ``plot holes''? 
Our work suggests these semantic commit interfaces should heavily prioritize recall over precision, as a missed conflict across a 100k+ word novel may be catastrophic, compared to lightly reviewing false positives.

\balance

\bibliographystyle{ACM-Reference-Format}
\bibliography{sample-base}

\appendix

\section{Conflict classification prompt}\label{app:conflict-classificaton-prompt}

\lstset{
  basicstyle=\ttfamily\small,
  backgroundcolor=\color{gray!10},
  breaklines=true,
  frame=single,
  columns=fullflexible,
  breakindent=0pt,
}

The following system prompt to GPT-4o is used for \textsc{SemanticCommit}'s NLI-like conflict classification into one of three categories: yes (contradiction), no (neutral), and ambiguous. It primes the classification by first asking the model to provide a reason for its classification before giving it. We deliberately do not include few-shot examples as, while such a method works well on limited benchmarks, it risks overfitting in unknown ways to example data, and we wanted to leverage the general knowledge of the LLM as much as possible. We iterated over this prompt by loading it into ChainForge and running it across the Mars and Labyrinth benchmarks to iterate with confidence. We erred ultimately on the side of caution to enhance recall---with minor changes to this prompt, for instance, detection immediately becomes more targeted, at the expense of less ``ambiguous'' markers. We do not claim this prompt is the best for this task and model, just that it was sufficient for our prototype.

\begin{lstlisting}
You are an information management system where the user has stored unstructured documents for a project that they are working on. All of these documents are important. The user has a new piece of information they wish to add to the project. Your job is to review the context and decide if the new piece of information impacts the information in the context.

For the new information to be considered impacted, it must directly contradict something in the context or the new information must be indirectly contradict something in the context. Note, you'll only be given partial context of the documents, ensure all the ways the new information could be related to the context are considered.

Format your response as a JSON object as follows:
{
  "reasoning": "Explain your reasoning why the new text is conflicting with the existing text. Be very specific."
  "is_conflicting": "yes", "no," or "ambiguous" // "yes" if the new information contradicts something in the context, "ambiguous" if the new information *might* indirectly contradict something in the context but it's hard to tell, and "no" if it is clear there is no contradiction
}
\end{lstlisting}

With this system prompt, each piece of existing information is compared with the new information via the input prompt: 

\begin{lstlisting}
Existing text:
{existing_info}

New text:
{new_info}
\end{lstlisting}

\section{Technical Evaluation} \label{appendix:experimental-eval}

In this section, we further detail our technical evaluation leading us to our chosen model \texttt{GPT-4o}. 
The selection was based on our analysis of conflict classification performance under varying false positive rates (FPRs) and specifically, as the input context includes an increasing number of non-conflicting chunks of information. We also made the decision based on the model latencies.

\begin{figure}[ht]
    \centering
    \includegraphics[width=0.45\textwidth]{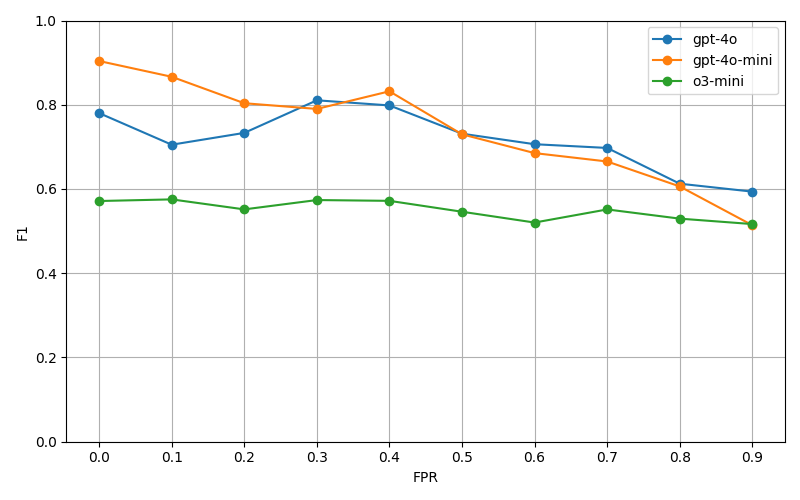}
    \caption{F1 score for conflict detection on the \emph{Labyrinth} of \texttt{GPT-4o}, \texttt{GPT-4o-mini}, and \texttt{o3-mini} on labyrinth at different false positive rates (FPR).}
    \label{fig:model-accuracy-fpr}
\end{figure}

\begin{figure}[ht]
    \centering
    \includegraphics[width=0.45\textwidth]{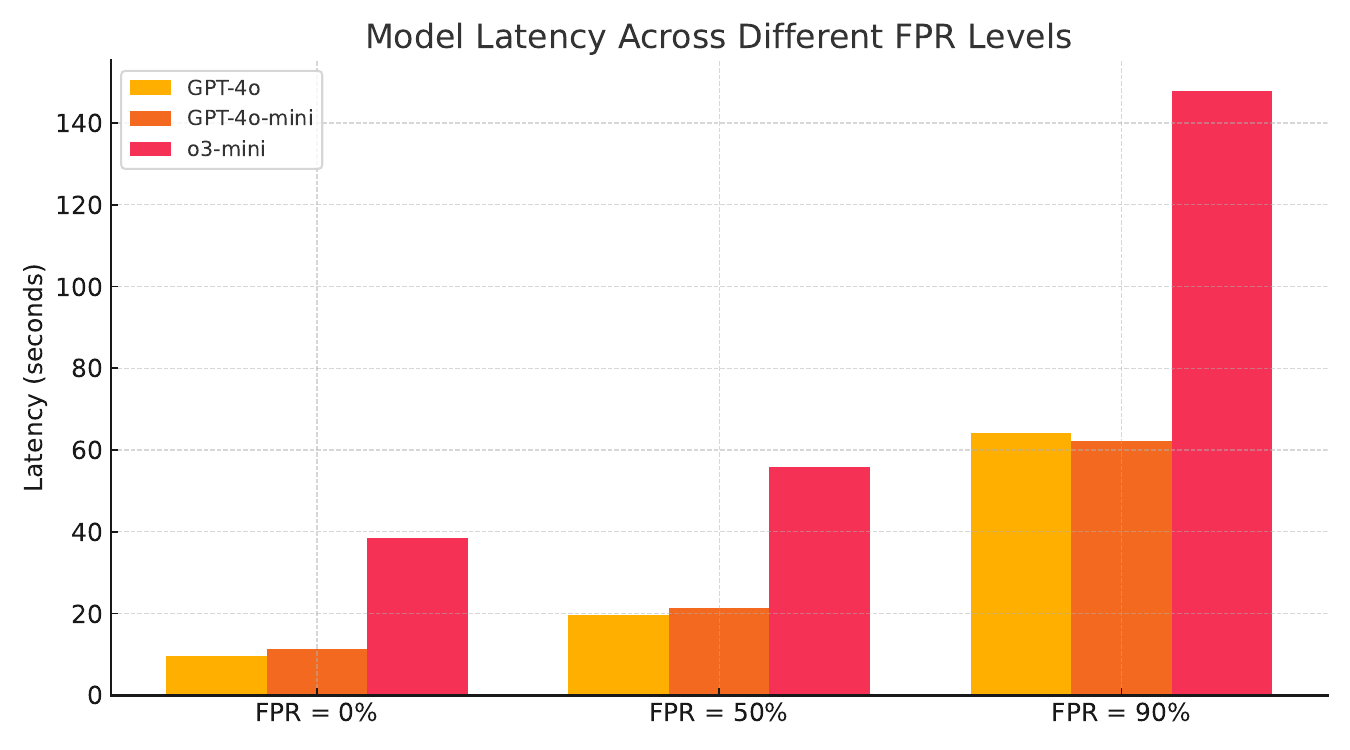}
    \caption{Average latency (in seconds) of \texttt{GPT-4o}, \texttt{GPT-4o-mini}, and \texttt{o3-mini} on labyrinth for conflict detection with different false positive rates (FPR); $0$, $0.5$, and $0.9$}
    \label{fig:model-latency}
\end{figure}

Figure~\ref{fig:model-accuracy-fpr} shows the F1 score for conflict detection on our \emph{Labyrinth} benchmark %
and how it changes as the false positive rate changes. 
We run this under idealized recall, i.e., we build an oracle for each edit with the set of actual conflicting documents presented to the \emph{conflict classification} stage and we inject additionally non-conflicting ones to reach a specific FPR.

We find that \texttt{GPT-4o} and \texttt{GPT-4o-mini} have comparable performance numbers. 
For both models, we see a decrease in their F1 score as the FPR increases. 
\texttt{3o-mini} on the other hand stays consistent in its F1 score as FPR increases but has a worse performance than both \texttt{GPT-4o} and \texttt{GPT-4o-mini}.

We also report the latency results in Figure~\ref{fig:model-latency} for the conflict classification stage of the pipeline. We observe that \texttt{GPT-4o} and \texttt{GPT-4o-mini} exhibit very similar latency while \texttt{o3-mini} is always more than two times slower. 
As a result, \texttt{GPT-4o} emerges as the model of choice for the conflict classification stage since it is broadly expected to be a more performant model. 

It is important to note that a fair evaluation of models on complex tasks such as conflict detection, is challenging due to the inherent ambiguity involved~\cite{jiang2022investigating}. This task required extensive prompt engineering and iterative refinement on our part. While the reported performance represents a best-effort evaluation, it is possible that the results for \texttt{o3-mini} reflect suboptimal prompting and do not fully capture the model's true capabilities on this task and benchmark. Nonetheless, even with comparable performance numbers, the model is relatively slow in its response time making it unsuitable for \textsc{SemanticCommit}, where low latency is essential for a good UX.

\section{Prompts for Baselines}\label{appendix:prompts-baselines}

In this section, we share the prompts we use for our two baselines \textsc{DropAllDocs} and \textsc{InkSync}. %

\subsection{DropAllDocs prompt}\label{appendix:dropalldocs-prompt}

Most of our toy benchmarks only consider adding new information (``add'' action). However, across our benchmarks there can be three possible actions: ``add'' (add new info), ``change'' (make a change), or ``edit'' (existing info) actions; in all cases, we are checking the chunks for conflicts. For ``edit'', we exclude the edited chunk ID. Below is the prompt template for DropAllDocs (which is fed in as an input prompt):

\begin{lstlisting}
You are an information management system where the user has stored unstructured documents for a project that they are working on. All of these documents are important and will be passed as a numbered list of statements, where each number acts as a unique identifier. {action_prompt} If there is a conflict, first explain the conflict ("CONFLICT:"), then provide a comma-separated list of ids which identify which specific statements that the new information conflicts with ("IDS:"). If there is no contradiction or ambiguity, return PASS. 

CONTEXT:
```
{all_docs}
```

NEW INFORMATION:
```
{new_info}
\end{lstlisting}

The ``action\_prompt'' depends on the action; for ``add'' and ``edit'' it is:

\begin{lstlisting}
The user has a new piece of information they wish to add to the project. Your job is to review the context and decide if the new piece of information conflicts or contradicts with any information in the context. The new information must directly contradict something in the context.
\end{lstlisting}

For ``change'' it is:

\begin{lstlisting}
The user has a change they wish to make to the project. Your job is to review the context and decide which exact items (statements) of the conflict need to be amended in order to make the change. The information must directly contradict something in the suggested change.
\end{lstlisting}

The ``all\_docs'' is \textit{a numbered list of texts}, i.e., the chunks in enumerated list form. 

\subsection{InkSync prompt}
\textsc{InkSync} generates suggested edits that contain the original text and suggested replacement text. 
We use the original prompt of InkSync~\cite{Labin2024Beyond} with modification for the conflict detection task. Similar to our approach in \textsc{DropAllDocs} and in order to have a fair comparison, we do not include few-shot examples. 
Below is the system prompt. We give InkSync all documents and an ``action\_prompt'', similar to \textsc{DropAllDocs} in Appendix~\ref{appendix:dropalldocs-prompt}. Note that we search for the original text across all documents and do not rely on \emph{document\_id} as part of the output format to avoid any possible \emph{document\_id} off-by-one errors in the LLM generation that we observed. 

\begin{lstlisting}
You are an information management system where the user has stored unstructured documents for a project that they are working on. All of these documents are important and will be passed as a numbered list of statements, where each number acts as a unique identifier. {action_prompt} For each document, if there is a conflict, suggest specific edits that could be implemented to resolve the conflict.

Output:
{{"edits": [
      {{"document_id": <document_id_1>, "original_text": <original_text_1>, "replace_text": <replace_text_1>}},
      {{"document_id": <document_id_2>, "original_text": <original_text_2>, "replace_text": <replace_text_2>}},
      ...
      ]
}}

Output Format:
- Only output the JSON object, do not output anything else.
- Follow the output format. Your entire output should be a valid JSON dictionary with the key: `edits`. The edits should be a list of valid edit objects, each with an `original_text`, `replace_text` keys. In each edit, the `original_text` text should be EXACTLY AS IS present in one of the numbered document at least as a sentence, otherwise the suggestion will be ignored.
- Your answer MUST START with: `{{"edits": "`
\end{lstlisting}

\section{User Study Tasks}

For completeness, we describe our study tasks here. From feedback in our pilot studies, we added a framing scenario for tasks in order to mimic the realistic scenarios that might appear in practice. This sacrificed a precise notion of ground truth---as we mention, conflicts are somewhat subjective and human annotators in NLI tasks frequently disagree---for the benefit of enhanced external validity.

\subsection{Task A - Integrate New Information into AI Memory of the User}

The participant is provided with the \textit{Financial Advice AI Agent Memory} intent specification loaded into the system, alongside a summary for easier onboarding.

They are also given a framing scenario: they are asked to imagine that they are an information management system that is managing memories about the user.

The participant is then asked to integrate the following new information, one at a time, using the provided system: 

\begin{enumerate}
    \item ``He’s decided to cut back on non-essential spending to start saving for a future home downpayment. Though it may take time, he feels it’s a necessary step toward greater financial independence.''
    \item ``Lately, he’s become more focused on investing, not just out of interest but as a strategic move to build up savings faster for a home downpayment. He’s begun reallocating some of his leisure time to research and portfolio management, seeing investment as a key part of his long-term plan.''
    \item ``He unexpectedly turned a 5 million won club fund into a significant profit through crypto futures trading, enough to afford a Porsche. This caught the attention of a school senior, who extended an unofficial offer to join a small proprietary trading firm.''
\end{enumerate}

\subsection{Task B - Update a Game Design Document}

The participant is provided with the \textit{Mars Game Design Document} intent specification loaded into the system, alongside a summary for easier onboarding.

They are also given a framing scenario: they are asked to imagine that they are a writer working with a game design team who just got done with an important team meeting on the game's direction, and is asked to update the game design document in response to changes decided at the meeting.

The participant is then asked to integrate the following new information, one at a time, using the provided system: 

\begin{enumerate}
    \item ``In the meeting, after reviewing player feedback, they decided to switch to 3D graphics with a fun claymation style to bring more warmth, charm, and personality to the world.''
    \item ``In the meeting, the team discussed adding more challenge and realism to colony management. Someone suggested a dynamic weather system on Mars, and the idea was approved to deepen gameplay—affecting energy production, plant growth, and encouraging strategic planning around unpredictable conditions.''
    \item ``In the meeting, the narrative team proposed shifting the setting from Mars to Venus to stand out from other space colony games. The team liked the idea—Venus’s extreme environment adds unique survival challenges and opens up fresh worldbuilding opportunities, so the change was approved.''
\end{enumerate}

\section{Design Explorations}

Our full project took place over the period of one year. For the interest of readers and transparency on our thought processes, we chronicle our design and architecture explorations here. 

\subsection{Design as a process of committing ideas}

We started our explorations by building brainstorming tool for game design where a human communicates with an AI agent. The agent stored mutual decisions in a shared game design document view, which the user could directly inspect and edit. The agent could then pass the document to another agent, to ask for code which implements the game. This brainstorming tool made us think through the process of design as a process of committing ideas at successive levels of importance and commitment on a project-specific document. We visualized this process as an idea passing through three overall stages that we term the \emph{idea space}, the \emph{paratext} (ideas and intent that the designer has committed to expressing), and the \emph{main text} (the actual implementation, or what the end-user of the design will see). One can analogize the paratext-main text relation to Hemingway’s iceberg theory in writing practice, where the bulk of the iceberg (the human intent) is hidden from the end-users of the implementation, with only parts of it directly represented (here, in the novel). We imagined that the user can interact with the AI agent to modify the paratext (e.g., conversation), or directly modify it (e.g., editing a document); either modification would then result in the AI system making required changes to the main text (and engaging the user in interactive dialogue when conflicts arise).

We quickly realized that, as interactions progress past the initial turns of chatting with the agent, the design document becomes lengthy and hard to navigate, with the AI's changes hard to verify and prone to errors: frequently over-revising or under-revising the document when integrating new information. Like other ``artifacts'' approaches by Anthropic and OpenAI, we were regenerating the entire document upon every suggested edit. This approach, dominant today, struggles to scale, is hard to verify (did something change?), and is too trusting, handing full agency to the AI to make edits immediately (and potentially change everything in the document) rather than engaging the user in a back-and-forth interaction. What happens when the user suggests an idea that is \textit{inconsistent} with an existing idea? Should the AI simply rewrite it? Or, more interactively, how can the AI raise this ``merge conflict'' to the user and engage them in an interactive resolution? And how can we help the user perform impact analysis---estimating the impact of integrating a new idea, with regards to the existing document---as project-specific documents become lengthy? 

\subsection{Architecture Design Iteration}
We set out to find a technical solution to this problem, and in the process began to better understand the space. We chronicle our explorations in prompt engineering and LLM-integrated system design here. 

Our first explorations were prompt prototyping to investigate the current capabilities of LLMs for semantic conflict resolution before proceeding. To perform our explorations in a structured way, we used the ChainForge open-source software, which helped us compare the performance of prompt templates across models, instruction variations and input data. We ran our tests over three contexts: docs about a game set on Mars, which was adapted from one author's personal game design document, and content from the 1986 design doc for the unpublished LucasArts game \textit{Labyrinth}, which we accessed via web archive and extracted from the report images using OCR. We chose this last document to provide a real game design doc that the LLM reasonably would not know about (compared to a published game), and for its longer document length (around 3 pages) that would stretch the limits of naive methods. We prototyped our initial design as a command-line interface (CLI) in Python, to reduce friction from UI development. 

We started prototyping with the most naive approach possible: giving the LLM a list of items in the prompt as context, and asking it to revise the items with respect to new information. We chose gpt-4o after comparing to gpt-3.5-turbo and gpt-4-turbo, as the latter two had trouble sticking to formatting constraints. After some initial prompt engineering, the approach overall worked well: the LLM seemed to revise what we expected, although it was sometimes more conservative with its revisions.

However, this naive approach did not scale as the context window grew large. Three issues stood out: as the context size increases, the LLM produced more errors and unnecessary rewordings or reorderings of information, as the LLM has to reproduce the entire context verbatim; it is costly, as the LLM needs to reproduce the entire context even for a small edit; and the document context may exceed the input token length for the LLM, requiring sharding to handle. Note that this method of asking the LLM to reproduce the entire project with one edit is what was being used in systems like OpenAI's Canvas interface or tldraw MakeReal ---doable for short documents, but fragile and not scale-able.

In our second iteration, we tried segmenting the context in a flatmap operation: map the changes over the context as separate LLM calls, then remerge them into one list. Though this approach sometimes led to reasonable edits, the LLM tended to mention the new information in too many places, producing duplicate information and unnecessary asides. This tendency seemed a limitation of the approach and not something that can be resolved with better prompt engineering: because each call presents a separate piece of context, the LLM has no awareness about how it's already adjusted the context, and this results in  changes that we perceive as redundant information when viewing the changes in aggregate. The approach also offered greater chances for the LLM to deviate from the expected output format, as it required many more calls. (We ultimately learned from this experience that we needed to separate conflict \textit{detection} from conflict \textit{resolution}---retrieval from generation---but did not know that here.)

We explored a compromise between these holistic and sharding solutions: to chunk the context into batches, using a metric such as embeddings or concept induction clustering, then feed the LLM these batches to revise (say, 5 short paragraphs) rather than single items. While this has the potential to suffer from some issues with our second method, it can hopefully reduce the amount of redundant information in the revised context and the potential for deviating from output format expectations (as less calls are made overall). As another alternative to random chunking, we also explored batching using LLooM~\cite{lam2024concept}, an open-source library that performs inductive coding on unstructured documents, clustering them by category. 

However, inductive clustering of information was also limited in practice. The issue is that it assumes a tree-like structure (a hierarchy of concepts), but in practice, there are many mutual dependencies that cross branches of the tree---say, a character description and the graphics to represent them. \textbf{Intent specifications have a \textit{heterarchical}, not hierarchical, structure.} A \textit{heterarchy} is a relation between elements of information that ``possess the potential for being ranked in a number of different ways'' \cite[p. 3]{crumley1995heterarchy}---here, information ranked and clustered just-in-time dependent on user query (the new information being integrated). Because this information could be separate in a hierarchical ranking, the LLM could not make consistent changes if one concept touches upon multiple clusters; and thus a similar problem of ``duplicate information during resolution'' again emerged. 

\begin{figure}
    \centering
    \includegraphics[width=\linewidth]{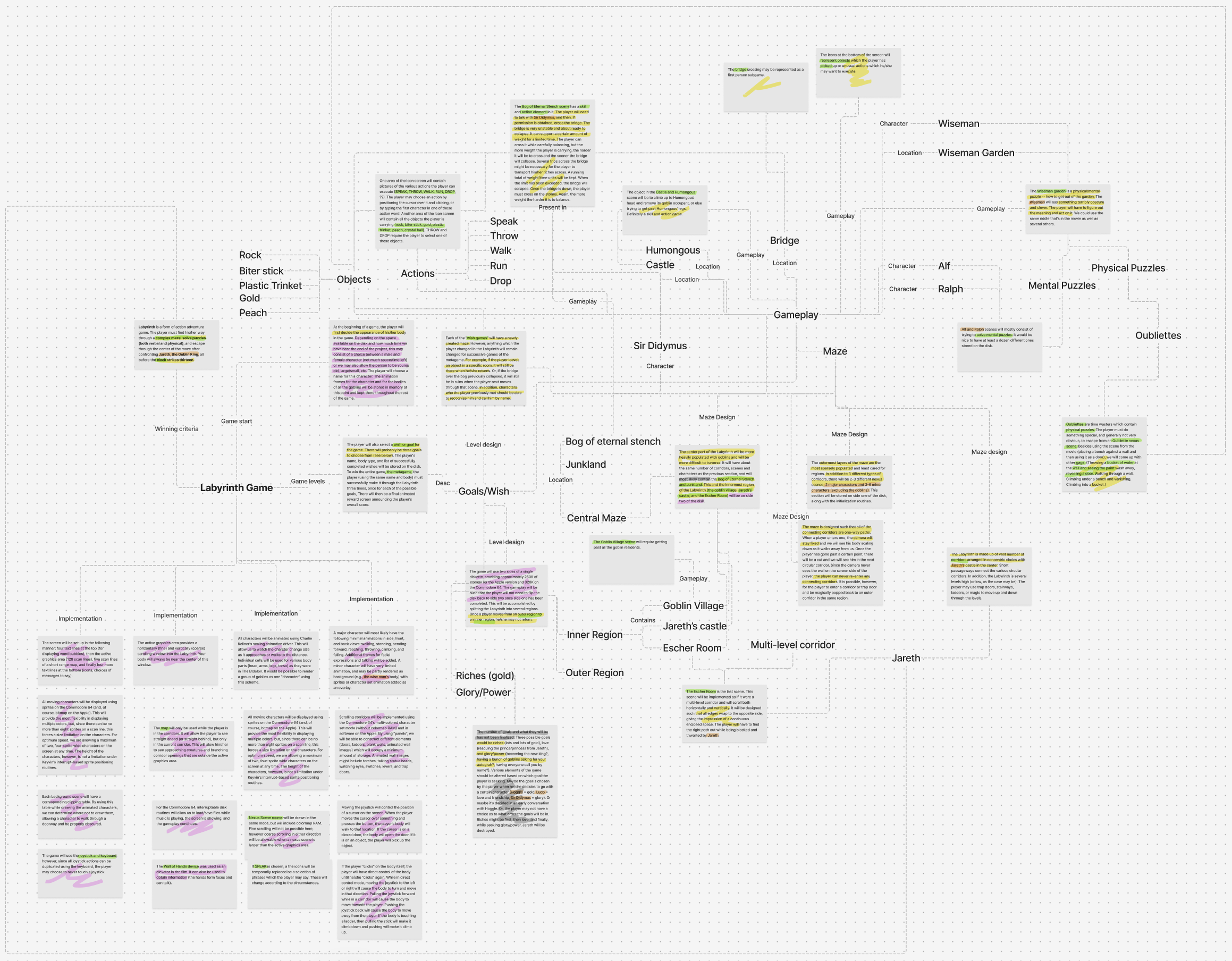}
    \caption{Creating a manual knowledge graph: Design explorations on how to represent the informational dependencies and entities of the Labyrinth game design document.}
    \label{fig:manual-kg}
\end{figure}

From here, we went back to the drawing board. %
To ground our explorations in concrete data and identify what was needed, one author took the notes from the Labyrinth design document---a game none of us were familiar with---put them on sticky notes in a digital canvas environment, and mapped how they were inter-related (Figure~\ref{fig:manual-kg}). They drew arrows between the ``dependencies'' and highlighted key entities and concepts. They also began to categorize types of information (e.g., levels, characters, suggestions). We realized that this dependency graph resembled a knowledge graph. (Note that our explorations of knowledge graphs at this point were right around the time the first KG-based solutions to RAG architectures were released---it was not a popular idea at the time.)

We learned a lot from this process. First, there are \textbf{``project-specific abstractions''---concepts and terms that have specific meaning to the project,} some of which \emph{are not capitalized} (and hence may not stand out upon first glance). Project-specific abstractions accumulate over time across interactions and become part of the unique \textit{common ground} shared between parties. An example of this in Labyrinth was the concept of a ``wish game.''  The ``wish game'' concept may not be recognized by a classical NER system as coherent, important term, since NER systems are trained on well-established rather than ad-hoc terminology. Another example is a term like ``labyrinth'' that has a meaning in English but may be re-defined or differently interpreted for the project context: here, Labyrinth stands for the game holistically, rather than a specific maze. Second, there are relations between chunks of information, that can be described in a form akin to a knowledge graph, e.g. ``is parent of,'' ``appears in,'' and which connect the pieces of information. Third, not all information is segmented correctly (e.g., one card might introduce both a new character and a monster type), emphasizing the importance of chunking during pre-processing. Fourth, real design documents also contain ``suggestions,'' such as remarks prefaced with a ``maybe'', that are not fully committed to the idea-space of the project, and which we may need to treat differently than fully committed information when resolving conflicts.\footnote{Reflecting, the author who had made the Mars game design doc also discovered that they had included suggestions in the doc, which hints that this may be behavior that occurs often in practice.}

We finally decided upon an LLM-induced knowledge graph architecture, implementing it ourselves independently before discovering HippoRAG~\cite{gutierrez2024hipporag} and adopting their Personalized PageRank technique. The induction step uses an LLM to generate entities and relations (nodes and edges), similar to the process from HippoRAG. The knowledge graph is updated upon confirmation of each resolved conflict (when user pressed the green checkmark). We found that this technique was effective at enhancing recall and extensible in that the retrieval step better scales when the information store grows large, shifting from (at best) a scan of all chunks of information $O(n)$, to a more limited filtering step after retrieval, $O(kg(n))$, which scales with the \textit{scope of conflict}---higher-degree changes, such as changing the setting of a game from Mars to Venus, could still result in latency, but lower-degree changes, like changing a specific detail of a character, are quite fast. In practice and real settings, users are liable to introduce small details much more often than vast changes of their intent---hence, KG-retrieval is quite optimal, since it is relatively low-cost and low-latency.\footnote{The caveat is that the KG needs to be initialized prior to use, which takes many LLM calls to accomplish, at least $O(n)$ for the initial number of chunks. However, this is a one-time, upfront cost.}

Our ``global conflict resolver'' prompt that is triggered when the user presses the global Make Changes button learned from the above experiences. For prototype purposes, we do in this step rely upon the LLM to perform a rewrite of the chunks. However, we only feed in the \textit{conflicting} chunks---the chunks our end-to-end pipeline has flagged as potentially conflicting. We then ask the model to rewrite the text. This was a bit fragile, and can be improved in later iterations; however, the key thing is that the model needs more context to not introduce repeat information when integrating the new information. Here is our ``Make Changes'' prompt:

\begin{lstlisting}
You are an information management system where the user has stored unstructured documents for a project that they are working on. The user has a new piece of information they wish {action_instructions} the project: "{newInfo}"

You have detected that this information conflicts or contradicts multiple existing pieces of information. Here are a list of existing texts. Edit the list as much as required to resolve the conflict. If you don't change some texts, that is OK. If you want to delete a text entirely, mark that line with "DELETE". Be as conservative as possible in your changes. Do not repeat information. Return the revised list of texts as a numbered list in Markdown format. Return only the numbered list.

TEXTS:
{all_docs}
\end{lstlisting}

\noindent where ``action\_type'' can be either "to add to" or "to integrate into." 

Note finally that semantic commits cannot be (fully) resolved using naive cosine similarity of vector embeddings. A naive approach may be to grab the most similar embeddings and only operate on those. Though this catches some semantic similarities, it only catches the most direct. For instance, consider the  statement: ``The enemy should sneak up behind the player when the enemy is not in the player's field of view,'' as part of the design document for a 3D game. If the user wants to change the game plan to target a 2D, top-down view, the semantic commit assistant should flag the enemy's behavior as potentially in conflict with the new design, requiring (at the very least) clarification. Embeddings alone would not reveal this hidden dependency between information. 

\section{Conflict Classification: Criteria—Direct, Ambiguous, and Non-Conflict in the Context of Possible Worlds and Reference Ambiguity} \label{appendix:conflict-classification}

This section describes our classification criteria for direct conflicts, ambiguous conflicts, and non-conflicts in further detail, incorporating essential background on issues such as possible worlds and reference ambiguity. Four coauthors developed this criteria over time in an iterative process while creating the benchmarks and mutually decided upon these definitions. We provide this information for readers who are looking to make similar benchmarks. %

\subsubsection{Direct Conflict}

A \textbf{direct conflict}, or \textit{contradiction} in NLI~\cite{jiang2022investigating}, refers to a case where two statements, when evaluated within the same possible world~\cite{sep-possible-worlds}—that is, under the same common ground and a “normal interpretation”—cannot both be true simultaneously. In other words, if the statements share the same context, the truth of one necessarily entails the falsity of the other, meaning that there is no possible world, under a normal interpretation, in which both statements coexist as true. This concept reflects the notion of the Law of Non-Contradiction in formal logic\cite{sep-contradiction}, and it includes situations where, even in the absence of explicit negation, people intuitively judge the statements as “this doesn’t make sense.” For example, “Alice was born in 1990” and “Alice is 20 years old in 2025” present a temporal contradiction—since they cannot both be true for the same Alice, they are classified as a direct conflict.

\subsubsection{Ambiguous Conflict}

An \textbf{ambiguous conflict} refers to a situation where two statements appear contradictory on the surface but can be reconciled when additional context or background knowledge is introduced within the framework of a “normal interpretation.” This is similar to Chen et al.'s notion of \textbf{reference ambiguity} \cite{chen2025onreference}. Under common assumptions or everyday interpretations, the two statements might seem to conflict; however, if there exists a reasonable possible world—within that normal interpretation—in which both statements can be simultaneously true due to specific circumstances or extra information, then they are considered to be in an ambiguous conflict. For example, the statements “Sally sold a boat to John” and “John sold a boat to Sally” generally appear mutually exclusive, but if each transaction pertains to a different boat, then there is a possible world in which both statements hold true. Similarly, consider “Alice’s father is a scientist” versus “Alice’s father is a police officer.” While these intuitively seem to conflict, if, for instance, Alice’s father is a forensic specialist within the police force or Alice has more than one parental figure that she calls father, then there exists a possible world in which both statements are valid. Consequently, such cases are interpreted as ambiguous conflicts.

\subsubsection{Non-conflict}
A \textbf{non-conflict}, or \textit{neutral} classification in NLI~\cite{jiang2022investigating}, refers to situations where two statements can coexist without incident, under most normal interpretations of the world shared between evaluators. 
We define the term ``\textbf{normal interpretation}'' as an interpretation carried out within the common ground that communicating parties share by ``common sense'' and within the broad society that they share. If two statements are evaluated within the same context or possible world, and if information or background knowledge strongly suggests the possibility that they can coexist, then they can be considered as not being in conflict. If, however, an interpretation completely deviates from the scope expected in the common world we share—such as when discussing Alice while presupposing that every Alice is a different entity, or when making claims like ``in this worldview 1+1=3'' or ``100 is less than 99'' without any context—then such cases would be classified as in conflict.

It is obviously true that these cannot be completely formal definitions: the scope of “normal interpretation” varies from person to person or situation to situation in terms of subjectivity and diversity. Some individuals might judge the same statements as conflicting, while others might not see any conflict. %
This acknowledges that the \textit{truthlikeness} (or \textit{verisimilitude}~\cite{sep-truthlikeness}) of a statement is evaluated relatively according to individual interpretation, and under the premise that completely measurable contradictions or absolute truths do not exist objectively, we can categorize statements into direct conflict, ambiguous conflict, and non-conflict based only on these relative criteria.

\subsubsection{Example of the Difficultly and Contextual Nature of Conflict Classification: The Warp Drive Example}

To help readers better understand the nuances of conflict classification, consider two chunks of information regarding a warp drive in a game. One chunk states that the warp drive can exceed the speed of light, while the other chunk presents a setting in which the warp drive ``first moves slowly and then suddenly operates at a very high speed.'' If a new chunk stating “no material can move faster than the speed of light” is added, then a contradiction naturally arises with the first chunk. However, in the case of the second chunk, the question of whether it exceeds the speed of light and the existence of a ``very high speed''---which in the full context implies, but does not clearly state, faster-than-light travel---are not necessarily contradictory. In this case, and given the full context and the commonsense association of a ``warp drive'' with FTL travel, the latter chunk \textit{may} be marked as in ambiguous conflict and flagged for human review, at which point the human could clarify the precise meaning of ``very high speed.'' However, it is also entirely possible to imagine a ``neutral'' classification here, and other information in the knowledge base might further attenuate our decision. Any benchmark of conflict classification, while trying to be as precise as possible and reach mutual agreement across coders, can still be subject to debate in such situations.

\end{document}